\documentclass[]{aa}  

\usepackage{graphicx}
\usepackage{txfonts}
\usepackage[dvipsnames]{xcolor}
\usepackage{subcaption}
\usepackage{float}
\usepackage{amsmath}
\usepackage[export]{adjustbox}

\newcommand{\GeV}{\text{GeV}}
\newcommand{\kpc}{\text{kpc}}
\newcommand{\cm}{\text{cm}}
\newcommand{\s}{\text{s}}
\newcommand{\G}{\text{G}}
\newcommand{\derivative}[2]{\frac{\mathrm{d} #1}{\mathrm{d}#2}}
\newcommand{\tderivative}[2]{\tfrac{\mathrm{d} #1}{\mathrm{d}#2}}

\newcommand{\muJybeam}{\mu {\rm Jy\, beam}^{-1}}

\newcommand{\parderivative}[2]{\frac{\partial #1}{\partial #2}}
\newcommand{\sparderivative}[2]{\frac{\partial^2 #1}{\partial #2 ^2}}

\newcommand{\cs}{\langle\sigma v\rangle}

\begin{document}

   \title{Weakly interacting massive particle cross section limits from LOFAR observations of dwarf spheroidal galaxies}

   \author{L.~Gajovi\'c\inst{1}
          \and
          F.~Welzmüller\inst{1}
          \and
          V.~Heesen\inst{1}
          \and 
          F.~de Gasperin\inst{2,1} 
          \and 
          M.~Vollmann\inst{3}
          \and
          M.~Br\"uggen\inst{1}
          \and
          A.~Basu\inst{4} 
          \and 
          R.~Beck\inst{5}
          \and 
          D.~J.~Schwarz\inst{6}
          \and 
          D. J.~Bomans\inst{7}
          \and 
          A.~Drabent\inst{4}
          }

   \institute{Hamburger Sternwarte, University of Hamburg, Gojenbergsweg 112, 21029 Hamburg, Germany 
   \and 
   INAF - Istituto di Radioastronomia, via P. Gobetti 101, 40129 Bologna, Italy
   \and
    Institut f\"ur Theoretische Physik,
Auf der Morgenstelle 14, Eberhard Karls Universit\"at T\"ubingen,
72076 T\"ubingen, Germany
    \and
    Th\"uringer Landessternwarte, Sternwarte 5, 07778 Tautenburg, Germany
    \and
    Max-Planck Institut für Radioastronomie, Auf dem H\"ugel 69, 53121 Bonn, Germany
    \and 
    Fakult\"at f\"ur Physik, Universit\"at Bielefeld, Postfach 100131, 33501 Bielefeld, Germany
    \and 
    Ruhr University Bochum, Faculty of Physics and Astronomy, Astronomical Institute, Universit\"atsstr. 150, 44780 Bochum, Germany
    }

   \date{Received 18 November 2022 / Accepted 19 March 2023}

 
  \abstract
   {
   Weakly interacting massive particles (WIMPs) can self-annihilate, thus  providing us with a way to indirectly detect dark matter (DM).
   Dwarf spheroidal (dSph) galaxies are excellent places to search for annihilation signals because they are rich in DM and background emission is low. If O(0.1--10$~\mu$G) magnetic fields in dSph galaxies exist, the particles produced in DM annihilation emit synchrotron radiation in the radio band.}  
   {We used the non-detection of 150 MHz radio continuum emission from dSph galaxies with the LOw Frequency ARray (LOFAR) to derive constraints on the annihilation cross section of WIMPs in electron--positron pairs. Our main underlying assumption is that the transport of the cosmic rays can be described by the diffusion approximation, which necessitates the existence of magnetic fields.
   }
   {We used observations of six dSph galaxies in the LOFAR Two-metre Sky Survey (LoTSS). The data were reimaged, and a radial profile was generated for each galaxy. We also used stacking to increase the sensitivity. In order to derive upper limits on the WIMP cross section, we injected fake Gaussian sources into the data, which were then detected with 2$\sigma$ significance in the radial profile. These sources represent the lowest emission we would have been able to detect. 
   }
   {We present limits from the observations of individual galaxies as well as from stacking. We explored the uncertainty due to the choice of diffusion and magnetic field parameters by constructing three different model scenarios: optimistic (OPT), intermediate (INT), and pessimistic (PES). Assuming monochromatic annihilation into electron--positron pairs, the limits from the INT scenario exclude thermal WIMPs ($\langle \sigma v \rangle\! \approx 2.2 \times 10^{-26}~\rm cm^3\,s^{-1}$) below 20 GeV, and the limits from the OPT scenario even exclude thermal WIMPs below 70 GeV. 
   The INT limits can compete with limits set by {\it Fermi}--LAT using $\gamma$-ray observations of multiple dwarf galaxies, and they are especially strong for low WIMP masses. }
   {}


   \maketitle
%
\section{Introduction}

Dark matter (DM) is known to interact gravitationally, and only weakly via other fundamental forces of nature, which makes it difficult to observe. Among the most promising candidates for DM are weakly interacting massive particles \citep[WIMPs;][]{jungman96}, (quantum chromodynamics)  axions \citep{peccei77a, peccei77b, weinberg78, wilczek78} or axion-like particles \citep{kim87,jaeckel10}, massive compact halo objects \citep{alcock00}, sterile neutrinos \citep{ibarra15}, primordial black holes \citep{hawking71}, and modification of the Newtonian dynamics as an alternative to explain the effect of DM without additional mass \citep{milgrom83}. WIMPs are particularly appealing candidates for DM and by far the most scrutinized.

In the WIMP hypothesis, the number density of DM particles freezes out when the expansion rate of the Universe becomes higher than their annihilation rate, so the number density of DM particles becomes constant. From this, the theoretical thermal relic annihilation cross section is calculated to be $\langle \sigma v \rangle\! \approx 2.2 \times 10^{-26}~\rm cm^3\,s^{-1}$ for WIMP masses above 10 GeV and predicted to increase at lower masses \citep{stiegman12}. There are also alternative production scenarios such as freeze-in, where DM never attains thermal equilibrium with the primordial plasma of standard model particles in the early Universe \citep{hall10}, as well as several scenarios where DM is in thermal equilibrium: hidden sector freeze-out \citep{douglas07, cheung11}, zombie DM \citep{kramer21}, pandemic DM \citep{bringmann21, hryczuk21}, and others.

There is ongoing research attempting to directly detect WIMPs, for example EDELWEISS \citep{sanglard05}, the XENON-1T experiment \citep{aprile17}, and CRESST \citep{abdelhameed19}. Another way to achieve a detection is an indirect astrophysical search targeting particles that were created by or interacted with DM. Searches using $\gamma$-ray observations of nearby dwarf spheroidal (dSph) galaxies using the {\it Fermi} Large Area Telescope \citep[{\it Fermi}--LAT;][]{ackermann15, hoof20} and the High Energy Stereoscopic System \citep[HESS;][]{aharonian08} are able to provide new stringent upper limits for the WIMP annihilation cross section that depend on the particle mass. Recent work by \citet{delos22} shows that if DM is thermally produced WIMPs, thousands of Earth-mass prompt cusps should be present in every solar mass of DM. 
This is manifested in a drastic increase in the expected DM annihilation signal compared to a smooth DM distribution, substantially tightening observational cross-section limits. The most recent $\gamma$-ray results \citep{hoof20}, assuming a smooth DM distribution, exclude the thermal relic annihilation cross section for WIMPs with masses of $\lesssim$ 100 GeV annihilating through the quark and $\tau$-lepton channels; and for WIMPs with masses of $\lesssim$ 20 GeV annihilating through the electron--positron channel.

Radio continuum observations provide a complementary approach to constraining the WIMP annihilation cross section. Such searches exploit the fact that highly energetic electron--positron pairs ($e^\pm$ pairs) produced by the annihilation of WIMPs give rise to synchrotron emission in the presence of magnetic fields \citep{colafrancesco07}. Although other WIMP annihilation channels can be explored, in this work we focus on WIMP self-annihilation into $e^\pm$ pairs.
For WIMP masses on the order of 10 GeV and microgauss ($\mu$G) magnetic fields, the critical frequency of synchrotron radiation, $\nu=3eBE^2/(2\pi m_e^3c^5),$ is on the order of a few gigahertz \citep{rybicki86}, which means the WIMP signal can only be detected at frequencies lower than that. Lowering the frequency below 1~GHz opens up the possibility for constraining WIMPs with masses down to 1~GeV.

Cosmic ray (CR) electrons and positrons produced via DM annihilation interact with the magnetic field and create an extended source that traces a diffusion halo due to the emitted synchrotron radiation. A common search strategy is to consider radial intensity profiles of the extended emission from these halos and compare them with modeled DM annihilation signals \citep{cook20,vollmann20}. \cite{vollmann21} presents semi-analytical formulae for these models, which were shown to be in reasonable agreement with the more sophisticated numerical methods \citep{regis15,regis21}. 

Dwarf spheroidal galaxies are very promising systems to look for DM-related emission since in these systems the radio continuum emission is almost uncontaminated by baryonic emission of primary CR electrons due to a low level of star formation \citep{colafrancesco07, heesen21}. Also, stellar-dynamical observations indicate that dSph galaxies are DM dominated \citep{geringer15}, meaning that the signal from WIMP annihilation is expected to be bright compared to other types of galaxies. Most known dSph galaxies are satellites of either the Milky Way or the Andromeda galaxy and are therefore close to Earth. This allows us to resolve the spectra of their stellar light well, and the estimates of their DM content are relatively precise \citep{hutten22}.

The LOw Frequency ARray \citep[LOFAR;][]{vanhaarlem13} is the ideal instrument for radio continuum searches of WIMP annihilation. LOFAR is a radio interferometer operating at low frequencies, from 10 to 240 MHz. LOFAR combines the high angular resolution needed to identify compact background sources with the high sensitivity to extended emission needed to detect the signal from WIMPs. A LOFAR high band antenna (HBA) search for DM in the dSph galaxy Canes Venatici I using such a strategy was already performed by \cite{vollmann20}. That proof-of-concept study shows that, for an individual galaxy, the limits are comparable to or even better than that from {\it Fermi}--LAT, assuming reasonable values for the magnetic field strength and diffusion coefficient. This work expands upon that study by considering six galaxies observed with LOFAR HBA. In order to improve the WIMP annihilation cross-section limits, we stacked the signal from these six galaxies in various ways. 
We thus present improved upper limits on the annihilation of low-mass WIMPs into $e^\pm$ pairs in the GeV--TeV mass range using new radio continuum data.

This work is organized as follows. In Sect.~\ref{se:dark_matter_calculations} we present our theoretical calculations, where we convert the presence of DM into a radio continuum signal. Section~\ref{s:methodology} presents our employed methodology, including how we dealt with the LOFAR radio continuum data and how we improved our signal-to-noise ratio via both profile and image stacking. Section~\ref{s:results} contains our results. We finish off in Sect.~\ref{s:disccusion_and_conclusions}, which contains the discussion and conclusions.

\section{Dark matter calculations}
\label{se:dark_matter_calculations}
\subsection{Cosmic ray injection}

The injection of standard model particles by WIMP annihilation is described by for example \citet{lisanti16}. The injection rate of CR electrons or positrons is expressed as

\begin{equation}
    s(r,E) = \frac{\rho^2(r)}{2m_\chi^2} \derivative{\cs}{E}\Bigg|_{\chi\chi \rightarrow e^+e^-},
\end{equation}
where $m_\chi$ is the WIMP mass, $E$ is the CR energy and $r$ is the galactocentric radius.
The injection rate also depends on the DM density profile $\rho(r)$ and the annihilation cross section $\cs$. We assume a Navarro--Frenk--White \citep[NFW;][]{navarro97} model for the density profile given by

\begin{equation}
    \rho_{\text{NFW}}(r) = \frac{\rho_s}{\frac{r}{r_s}\left(1 + \frac{r}{r_s}\right)^2}\ ,
    \label{eq:NFW_profile}
\end{equation}
where $\rho_s$ is the characteristic density and $r_s$ is a scale length. The velocity- and spin-averaged cross section for WIMP annihilation into $e^\pm$ pairs per unit energy are obtained as

\begin{equation}
    \derivative{\cs}{E} = \sum_{f^+f^-}\text{BR}_{f^+f^-}\cs\derivative{N_{f^+f^-\rightarrow e^+e^- + X}}{E} \ ,
    \label{eq:cross-section_derivative}
\end{equation}
where BR$_{f^+f^-}$ is the branching ratio that describes the weighting of the elements for any standard model particle pair $f^+f^-$, into which the WIMPs will annihilate. This is calculated with for example the Fortran package package \textsc{DarkSUSY} \citep{bringmann18}, but we assume annihilation into monochromatic electron--positron pairs. Then the derivative in Eq.~(\ref{eq:cross-section_derivative}) simplifies to $\tderivative{N_{f^+f^-\rightarrow e^+e^-+X}}{E}=\delta(E-m_\chi)$. Once an electron or positron is injected into the DM halo, it diffuses through the turbulent magnetic field in the halo of the dSph galaxy while emitting synchrotron radiation.

\subsection{Cosmic ray diffusion}

The electron mass is always much lower than the WIMP mass, $m_e \ll m_\chi$, so the $e^\pm$ pairs are ultra-relativistic. Since magnetic fields in galaxies are mostly turbulent \citep{beck15}, we assume the propagation of $e^\pm$ pairs to be dominated by diffusion \citep{colafrancesco07,regis14}. Hence, the CR propagation is described with the stationary diffusion--loss equation, which we adopt as our transport model. Due to the spherical geometry of dSph galaxies, we further assume an isotropic momentum distribution, such that the diffusion--loss equation depends only on galactocentric radius, $r$, and CR energy:
\begin{equation}
    D(E)\frac{1}{r} \sparderivative{}{r}\left[n_e(r,E)r\right] + \parderivative{}{E}\left[b(E,B)n_e(r,E)\right] + s(r,E) = 0\ .
    \label{eq:transport_model}
\end{equation}
The parameter $b(E,B)$ describes the total energy loss-rate and $D(E)$ describes the diffusion coefficient. As a boundary condition, we assume that the CR number density (per unit volume and energy) $n_e$ vanishes at diffusion-halo radius $r_h$, as the diffusion coefficient rises to infinity due to vanishing magnetic fields, so that $n_e(r_h,E) = 0$. We assume that $r_{\rm h}$ is on the order of the half-light radius $r_\star$. These assumptions, including stationarity, are well justified and became the de facto standard over the years \citep{colafrancesco06, mcdaniel17}. 

Contributions to the total energy loss include synchrotron radiation, $b(E,B)_{\rm sync}$,  and inverse Compton scattering, $b(E,B)_{\rm ICS}$, losses. Other energy losses, such as bremsstrahlung and ionization losses, are suppressed due to the low density of both ionized and neutral gas \citep{regis15}. The total energy loss of ultra-relativistic CR $e^\pm$ in nearby galaxies (at redshift $z=0$) is then
\begin{multline}
    b(E,B) = b_{\rm sync}(E,B)+b_{\rm ICS}(E,B) \\
    \approx 2.546\times 10^{-17}\left[1 + \left(\frac{B}{3.24~\mu\G}\right)^2\right]\left[\frac{E}{1\GeV}\right]^2\GeV\,\rm s^{-1} ,
\end{multline}
where $B$ is the magnetic field and 3.24 $\mu$G is the field with the energy density of the cosmic microwave background (CMB) photons at z=0. Inverse Compton scattering on radiation fields other than the CMB can be neglected. This is because the radiation energy density due to stellar light, $u_\star$, is quite low in comparison to that of the CMB, $u_{\rm CMB}$, with $u_\star/u_{\text{CMB}} \approx 0.3\,\% \left(L_V/10^6 L_\odot\right)\left(r_\star/\kpc\right)^{-2}$ , where $L_V$ is the $V$-band luminosity \citep{vollmann21}. 

The average magnetic field strength in dSph galaxies is generally poorly known. 
In order to observe any radio continuum emission, the CR electrons have to be confined to the plasma, which implies that the magnetic field energy density cannot be too low in comparison with the CR energy density, even if the exact relation is not clear. For our typical sensitivities of 0.25-0.5 mJy\,beam$^{-1}$, frequency $\nu= 144$ MHz, beam size 20", spectral index $\alpha= -0.8$, path length $l= 400$ pc, and proton/electron ratio $K_0= 0 $, an equipartition magnetic field strength of $\approx$2\, $\mu$G is calculated for an $e^\pm$-plasma \citep{beck05}. We adopted a more conservative value of 1\,$\mu$G here, but varied it by an order of magnitude to account for the large uncertainties.

For the diffusion coefficient, we assume an energy-dependent power law,
\begin{equation}
    D(E) = D_0\left(\frac{E}{1\:\GeV}\right)^\delta\ ,
    \label{eq:diffusion_coeff}
\end{equation}
where $D_0 = D(1\,\GeV)$ and the power-law index, $\delta$, describes the energy dependence, which is determined by the adopted turbulence model. The diffusion coefficient and its energy dependence in dSph galaxies are prone to uncertainties. We adopted a value of $10^{27}~\rm cm^2\,s^{-1}$ for $D_0$, which is in agreement with observations of nearby dwarf irregular systems \citep{murphy12,heesen18}. As no measurements exist for diffusion in dSph galaxies, we varied $D_0$ within two orders of magnitude. For the energy dependence, we assumed Kolmogorov-like turbulence resulting in $\delta = \tfrac{1}{3}$ \citep{kolmogorov91}. This model is supported by observations of the Milky Way \citep{korsmeier16}. A key question is to what extent there is turbulence in the magnetic field structure, if it is present at all. Observations of the Milky Way and other galaxies usually fall into the two categories. Either the turbulence is extrinsically generated, where turbulence cascades down from large scales to the small scales relevant for CR scattering; or the turbulence is intrinsically generated by the CRs themselves via plasma instabilities. The latter scenario is usually referred to a self-confinement \citep{zweibel13}. Because in dSph galaxies there is presumably no external source of turbulence such as supernova remnants, the self-confinement scenario is probably more appropriate. In star-forming galaxies this is likely the case as well for CRs with energy of a few GeV, where observations indicate a weaker energy dependence for CRs with less than 10 GeV energy \citep{heesen21b}. This is hence another source of uncertainty, at least for low WIMP masses.

\subsection{Model scenarios}
\label{ss:diffusion_scenario}

In order to deal with the uncertainties in the diffusion coefficient, $D_0$, and the magnetic field strength, $B$, we employed three different model scenarios. Our standard values ($D_0 = 10^{27}\:\cm^2\:\s^{-1}, B = 1\:\mu\G$) define the "intermediate" (INT) scenario. In the "optimistic" (OPT) scenario, we chose values that boost the DM signal. With a diffusion coefficient of $10^{26}\:\cm^2\:\s^{-1}$ and an average magnetic field strength of $10\:\mu\G$, the CRs diffuse slowly and emit more synchrotron radiation due to the strong magnetic field. We note that these values are highly optimistic and probably unrealistic, but they serve us as a reference point for the maximum signal we might possibly expect. Such a high magnetic field strength is only observed in regions of concentrated star formation in nearby dwarf irregular galaxies \citep{hindson_18a}. In the "pessimistic" (PES) scenario, we used a high diffusion coefficient of $10^{29}\:\cm^2\:\s^{-1}$ in a comparably weak magnetic field of $0.1~\mu\rm G$. In this situation, the CRs emit less synchrotron radiation and the DM signal is lower. For such weak fields, most of the CR electron energy, whether primary or secondary, would be lost via inverse-Compton radiation.

This study does not cover the case in which the magnetic-field strengths are even weaker than $\mathcal O(0.1~\mu$G). In this case, a ballistic description for the CR electron propagation would be more appropriate. Since we cannot exclude this possibility theoretically nor observationally, its study is left for future work.

\subsection{Diffusion regimes}

Because the full solution of the diffusion--loss equation (Eq.~\ref{eq:transport_model}) is rather complicated, \citet{vollmann21} defines three regimes, which allow one to simplify the solution. These regimes depend on the ratio of the CR diffusion timescale to the energy loss timescale. The diffusion timescale is
\begin{equation}
    \tau_{\text{diff}} = \frac{r_h^2}{D(E)}\ ,
    \label{eq:diffusion_time_scale}
\end{equation}
where $r_h$ is again the radius of the diffusion halo. As already mentioned, we assume that $r_{\rm h}$ is on the order of the half-light radius, $r_\star$.

The loss timescale from synchrotron radiation and inverse Compton scattering of CRs is \begin{equation}
    \tau_{\text{loss}} = \frac{E}{b(E,B)} = 1.245\left(1+\left(\frac{B}{3.24\:\mu \rm G}\right)^2\right)^{-1}\left(\frac{E}{1\:\GeV}\right)^{-1}\mathrm{Gyr}\ .
    \label{eq:energy_loss_time_scale}
\end{equation}
When $\tau_{\text{diff}} \gg \tau_{\text{loss}}$, one can assume that the CRs lose all their energy so rapidly that diffusion can be neglected and the first term in Eq.~\eqref{eq:transport_model} vanishes. We refer to this assumption as "regime $A$" or the no-diffusion approximation. "Regime $B$" is defined such that $\tau_{\rm diff}\approx\tau_{\rm loss}$. In this regime the we have to consider the full solution of Eq.~\eqref{eq:transport_model}. \citet{vollmann21} shows that the solution can be expressed as the sum of Fourier-like modes as a function of radius, where we consider only the leading term in this expansion. For $\tau_{\text{diff}} \ll \tau_{\text{loss}}$, one can neglect the second term of Eq.~\eqref{eq:transport_model}, as the CRs diffuse so rapidly that they leave the dSph galaxy without losing energy. We refer to this assumption as as "regime $C$" or the rapid-diffusion approximation.

\subsection{Synchrotron signal occurrence}

The radio continuum intensity, $I_{\nu}$, is the integral of the radio emissivity, $j_\nu(r)$, along the line of sight (LoS):
\begin{equation}
    I_{\nu} = \int_{\text{LoS}}dl\: j_\nu(r(l))\ .
\end{equation}
Following \citet{vollmann21}, the emissivity can be separated into a halo part, $H(r)$, a spectral part, $X(\nu)$, and a normalizing pre-factor,
\begin{equation}
    j_\nu(r) = \frac{\cs}{8\pi m_\chi^2 }\ H(r)\ X(\nu)\ .
    \label{eq:radio_emissivity}
\end{equation}
Both the halo and the spectral part depend on the diffusion regime. The regime-specific equations for the halo factors are presented in \citet{vollmann21}. For the halo part, we employ the leading-mode approximation (regime $B$). 
For the NFW profile (Eq.~\ref{eq:NFW_profile}) the halo function is 
\begin{equation}
    H_B(r) = h_B \frac{1}{r} \sin{\left(\frac{\pi r}{r_h}\right)} ,
    \label{eq:halo_part}
\end{equation}
where $h_B$ is the halo factor in units of $\rm GeV^2\,cm^{-5}$, which contains the part of $H_B$ that is independent of radius:
\begin{equation}
    h_B = 2\left[\text{si}\left(\pi\right) - \frac{8r_h}{\pi r_s} + \dots\right]\frac{\rho_s^2 r_s^2}{r_h}\ \ \text{where}\ \  \text{si}\left(x\right)=\!\int_0^x dt\: \tfrac{\sin\left(t\right)}{t}\ .
    \label{eq:halo_factor}
\end{equation}
It should be noted that Eq.~\eqref{eq:halo_part} is a simplified version only valid for emissivities. In order to compare with our measured intensities, we implemented the actual halo factor as calculated for intensities \citep[see][Appendix~B]{vollmann21}.

For the spectral part, it is not viable to only consider one regime, as $X(\nu)$ strongly depends on the environment, for example which energy loss mechanism is dominant in the specific situation. Hence, we used all three spectral parts for the regimes $A$, $B$, and $C$, respectively, from \citet{vollmann21}:
\begin{equation}
    X_A(\nu) =  \frac{2\sqrt{3}e^3 B}{m_e}\int dz\: \frac{F(z)}{z}\frac{E(\nu/z)}{b(E(\nu/z))}\int_{E(\nu/z)}^\infty dE\: S(E)\ , \label{eq:spectral_part_A}
\end{equation}
\begin{eqnarray}
    X_B(\nu) & = & \frac{2\sqrt{3}e^3 B}{m_e}\int dz\: \frac{F(z)}{z}\frac{E(\nu/z)}{b(E(\nu/z))}e^{-\eta(E(\nu/z))} \nonumber\\
    & & \int_{E(\nu/z)}^\infty dE\: S(E)e^{\eta(E)}\ , \label{eq:spectral_part_B}
\end{eqnarray}
\begin{equation}
    X_C(\nu) = \frac{2\sqrt{3}e^3 B}{m_e} \frac{r_h^2}{\pi^2}\int dz\: \frac{F(z)}{z}\frac{E(\nu/z)S(E(\nu/z))}{D(E(\nu/z))}\ ;
    \label{eq:spectral_part_C}
\end{equation} with $\eta (E) = \tfrac{\pi^2}{r_h^2}\int_E^\infty dE'\: \tfrac{D(E')}{b(E')}$. The function $F(x)$ is described by \citet{ghisellini88} as
\begin{equation}
    F(x) = x^2\left[K_{\frac{4}{3}}(x)K_{\frac{1}{3}}(x) - x \frac{3}{5}\left(K^2_{\frac{4}{3}}(x) - K^2_{\frac{1}{3}}(x)\right)\right]\ ,
\end{equation}
where $K_i(x)$ is the modified Bessel functions of the second kind. For monochromatic $e^\pm$ injection, the CR energy is given by $E(\nu) = \sqrt{2\pi m_e^3\nu/(3eB)}$ and the spectral injection function is $S(E) = \delta(E - m_\chi)$. All three spectral part formulae are related, where $X_A$ and $X_C$ are the limits for $X_B$ when assuming $\eta\rightarrow 0$ and $\eta\rightarrow \infty$, respectively.
 
Assuming that the total radio emissivity is due to DM annihilation, it is straightforward to see that the shape of the radial profile is determined only by the halo function $H(r)$ of Eq.~\eqref{eq:radio_emissivity}.
We can therefore express the emissivity as\begin{equation}
    j_\nu (r) = N_B\cdot  H_B(r)\ ,
\end{equation}
where $N_B$ is referred to as the signal-strength parameter that contains all the terms that do not depend on the radius \citep{vollmann21}. Now we get an expression for the cross section in terms of $N_B$:
\begin{equation}
    \cs = \frac{8\pi m_\chi^2 N_B}{X_j(\nu)},
    \label{eq:cross-section}
\end{equation}
where $j\in \lbrace A,B,C \rbrace$ specifies the diffusion regime. A similar approach is used in \citet{regis14} and \citet{vollmann20}. The factor $N_B$ connects predictions to observations, as it is proportional to the intensity of the DM signal in the radio band. 

\section{Methodology}
\label{s:methodology}
\subsection{LoTSS observations}

 The data used for our analysis are observed as part of the LOFAR Two-metre Sky Survey \citep[LoTSS;][]{shimwell17,shimwell19} and published in the second data release \citep[LoTSS-DR2;][]{shimwell22}. LoTSS is a deep low-frequency survey with LOFAR HBA at 144~MHz with 24~MHz bandwidth. LoTSS-DR2 includes observations of 841 pointings and covers 5634 square degrees of the northern hemisphere. LoTSS data have a maximum angular resolution of 6", referred to as high-resolution data, and additional low-resolution data at 20" angular resolution. The high-resolution data were important for the subtraction of point-like sources, whereas the low-resolution data allowed us to detect extended emission at high signal-to-noise ratios. The rms noise at 20" resolution is 50--100~$\mu\rm Jy\,beam^{-1}$ \citep{shimwell22}. 

We analyzed six dSph galaxies that are observed in the LoTSS-DR2. These galaxies have half-light radii between 20 and 600 pc with distances between 30 and 218 kpc (see Table~\ref{tab:galaxy_samples}). These galaxies are Canes Venatici I (CVnI), Ursa Major I (UMaI), Ursa Major II (UMaII), Ursa Minor (UMi), Willman I (WilI), and Canes Venatici II (CVnII).\begin{table*}[ht]
    \caption{Properties of the galaxies in our sample from LoTSS-DR2.}
    \centering
    \begin{tabular}{lccccccccc}
     \hline \hline
        dSph & R.A. (J2000) & Decl. (J2000) & $r_\star$ & $d$ & $r_s$ & $\rho_s$ & rms-noise & References\\
              & $[\rm h\:m\:s]$ & $[^\circ \: '\:"]$ & [pc] & [kpc] & [kpc] & [GeV cm$^{-3}$] & [$\mu$Jy beam$^{-1}$] &\\
        \hline
        \noalign{\smallskip}
        CVnI & 13:28:03.5 & +33:33:21 & 564 & 218 & 2.27 & 0.5186 & 115 & 1 \\
        UMaI & 10:34:52.8 & +51:55:12 & 319 & 97 & 3.20 & 0.5473 & 74 & 1 \\
        UMaII & 08:51:30.0 & +63:07:48 & 149 & 32 & 4.28 & 2.794 & 60 & 1 \\
        UMi & 15:09:08.5 & +67:13:21 & 181 & 76 & 0.394 & 12.10 & 103 & 1 \\
        WilI & 10:49:23.0 & +51:01:20 & 21 & 38 & 0.173 & 15.18 & 83 & 2, 3, 4\\
        CVnII & 12:57:10.0 & +34:19:15 & 74 & 160 & 8.04 & 1.331 & 68 & 1\\
        \hline
    \end{tabular}
    \tablebib{(1)~\citet{geringer15}; (2) \citet{martin08}; (3) \citet{willman11}; (4) \citet{sanchez-conde11}.
}
    \label{tab:galaxy_samples}
\end{table*} These are the only non-disturbed dSph galaxies observed with the LOFAR HBA at this point in time. Additional four galaxies with declinations above +20$^\circ$ \citep{geringer15} could be observed in the future. LOFAR sensitivity is greatly reduced at lower declinations \citep{shimwell17}, but with longer observational times it is possible to observe additional six galaxies with declinations between +10$^\circ$ and +20$^\circ$ \citep{geringer15,ackermann15}.

\subsection{Reimaging the LoTSS data}\label{ch:data_calibration}

We use recalibrated LoTSS data, where the calibration was specially tailored to our dSph galaxies \citep{vanweeren21}. We reimaged the $(u,v)$ data with {\sc WSClean v2.9} \citep{offringa14,offringa17}. The points in the $(u,v)$-plane were weighted using Briggs robust weighting as a compromise between uniform and natural weighting. A robustness parameter of {\tt robust=$-$0.2} was found to produce the highest signal-to-noise ratio for the extended emission on the scales that we are interested in. Further imaging parameters are listed in Appendix~\ref{a:re-calibration_parameters}.

We excluded emission on large angular scales ($\gtrsim$$1^\circ$) that can be attributed to the Milky Way \citep{erceg22} by excluding $(u,v)$ data at short baselines. We used lower limits to the $(u,v)$ range between 60 and 400~$\lambda$, corresponding to angular scales from to 7\arcmin\ to 46\arcmin, making sure that these scales are not smaller than the size of the galaxy. Compact sources were subtracted from the $(u,v)$ data prior to imaging. This was done by first producing a source catalog with the Python Blob Detector and Source Finder \citep[{\small PyBDSF};][]{rafferty19}. Since not all background sources have a point-like nature, we additionally used the "\`a trous wavelet decomposition module" integrated in {\small PyBDSF}. This module decomposes the residual maps resulting from the internal subtraction of the fitted Gaussians into wavelets of different scales \citep[see][]{holschneider89}. We used between two and five wavelet scales depending on the galaxy. Sources were then subtracted as Gaussians from the $(u,v)$ data using the Default Pre-Processing Pipeline software \citep[{\small DPPP};][]{vandiepen18}. 
 
The maps were deconvolved with the multi-scale and auto-masking options to remove any residuals comparable to the size of the galaxies. Maps were then restored with a Gaussian beam at 20\arcsec\ angular resolution. The reimaging steps (as well as further steps in the cross-section limits calculation) were automated using {\sc Python}.\footnote{\url{https://github.com/FinnWelzmueller/wimpsSoftware}}

\subsection{Calculation of the cross-section upper limits} \label{ch:limit_calculation}

To constrain the annihilation cross section of WIMPs, we used central amplitudes of the radial intensity profiles. To generate the radial profiles, we averaged intensities within annuli of increasing radius and constant width. For every galaxy, the width of an annulus was set to 20", which is equal to the full width at half maximum (FWHM) of the restoring Gaussian beam.

The expected shape of the radial profile is described by the halo function (Eq.~\ref{eq:halo_part}). To analyze the observed radial profiles we approximate the shape as a fixed-width Gaussian with a FWHM equal to $r_\star$. We note that the halo function depends on the density squared so the FWHM should indeed be approximated with $r_\star$ instead of 2$r_\star$, which we would expect from the density distribution. We varied the Gaussian central amplitude to best fit the observed radial profile. Since the size of the dSph galaxy diffusion radius is mostly unknown, it is important to verify the non-detection of a DM-related signals on various scales, not only the one we assumed earlier. Hence, we varied the FWHM of the Gaussians using values that are higher and lower than $r_\star$.

To mimic a DM halo, we injected fake sources directly into the point-source subtracted $(u,v)$ data. The fake source was constructed as a two-dimensional Gaussian with the FWHM equal to the stellar radius of the galaxy. This is a simplification as the real signal may be of a different shape. The simplification, however, has only a negligible effect on our inferred limits. The amplitude of the Gaussian was varied until we got a 2$\sigma$ detection by fitting a Gaussian to the radial profile, as for the purely observational profile. Since the Gaussian FWHM was fixed for each dSph galaxy, the only free parameter was the central amplitude, $a$, which is related to the factor $N_B$ in Eq.~\eqref{eq:cross-section}. The transformation between the central amplitude of the Gaussian radial intensity distribution, $a$, and the factor $N_B$ was done using the halo factor (Eq.~\ref{eq:halo_factor}): 

\begin{equation}
    N_B = \frac{\pi\left(0.4\right)^2}{2 h_B}\ a\ ,
    \label{eq:strength-giving_factor}
\end{equation}
where the numerical factors account for the different shapes of the Gaussian source fitted to the data and assumed form of the halo function described by Eq.~\eqref{eq:halo_part}. Specifically, the width $w$ (equivalent to the standard deviation) of the Gaussian is equal to $w=r_\star/(0.4d)$. Additionally, the spectral function approximation was calculated for the appropriate scenario. The upper limits on the cross section were determined as a function of WIMP mass by inserting $N_B$ into Eq.~\eqref{eq:cross-section}.

\subsection{Stacking}

In addition to looking at each galaxy separately, we combined the data from all galaxies through stacking. We used two different approaches for stacking the data. The first was to generate the radial profile for each galaxy separately, then to rescale to the stellar radius, and then stack the profiles. The second approach was to rescale and stack the images, and only then generate the radial profile from the stacked image.

\subsubsection{Stacking radial profiles}

Our first approach to stacking the data was to stack the radial intensity profiles that are rescaled to the stellar radius. We used the stellar radius instead of the NFW scaling radius as it is the much more reliable observable. We set the width of the annuli in which the radio intensity is averaged to 0.05$r_\star$ for each galaxy. This was larger than the beam FWHM for most, but not all, galaxies, which might cause a slight correlation between adjacent data points. Each data point in the radial profile was expressed in terms of $r_\star$ and the intervals between them were equal even if the actual size on the sky is different. We note that the intensities did not have to be corrected for distance. The rescaled radial profiles were combined by calculating the noise-weighted mean. By fitting a Gaussian to the stacked profiles we confirmed they are consistent with zero. 

After that we needed to handle the fake sources to be able to calculate the limits on the cross section. We first stacked the profiles with same fake sources as for the individual galaxies. The significance of the detection was higher than 2$\sigma$ so in the stacked profile we could detect a fainter DM signal. To determine exactly how much fainter, we lowered the intensity of all the injected sources by a common factor, so the flux density ratio between the galaxies remained the same as in the individual analysis. This factor was chosen to achieve a detection with a significance of 2$\sigma$ in the stacked radial profile. 

Once we had the stacked radial intensity profile for all galaxies, we repeated the fitting to determine the combined value for the Gaussian amplitude, $a$. To calculate limits on the cross section from the combined data of our galaxies, we used this value for $a$ and averaged all the other terms in Eq.~\eqref{eq:cross-section} for each WIMP mass and each scenario. This is justified because by averaging the other terms we calculated the average emission from galaxies and this is exactly what we got when stacking. 

\subsubsection{Stacking images}

Our second approach to stacking was to combine the galaxies in image space. 
This was done in the following steps:

\begin{enumerate}
    
    \item A rescaled cutout image of every galaxy was created with a size of 4$r_\star$ and 1367$\times$1367 pixel$^2$. The dimension in pixels was the median of all galaxies while ensuring there were at least seven pixels per beam FWHM. After the rescaling process, a single pixel sampled a larger section of the sky for galaxies with a larger angular diameter, compared to those with a small angular diameter, but this was a necessary compromise that we need to make for the image stacking. After this procedure, diffuse sources with equal flux density would appear identical, regardless of the distance and the stellar radius of the galaxy. 
    
    \item  The flux density variance $\sigma$ was calculated inside an annulus with an inner radius of $r_\star$ and an outer radius of $2r_\star$. 
    
    \item All images were stacked using the weighted mean. The weight was adopted as the inverse square variance (1/$\sigma^2$) of each image, so that galaxy images with lower noise contributed more to the stacked image. Cosmological surface brightness dimming is negligible because the galaxies are in the Local Group so we have not applied any weighting with redshift. 
    
\end{enumerate} 

From the final stacked image we generated the radial intensity profile (using the same algorithm as for the individual profiles) and confirmed a non-detection. To calculate the limits on the cross-section, we followed the same procedure as for the profile stacking. We adjusted the multiplication factors of injected sources and obtained the Gaussian amplitude, $a$. This combined value for $a$ was used in the calculation for the cross-section limits. Other necessary parameters that depend on galaxy properties were averaged.

\section{Results}
\label{s:results}

Our presentation of the results is split into three parts. We start with individual limits on the WIMP annihilation cross section (Sect.~\ref{ch:individual_limits}). In Sect.~\ref{ch:stacked_limits} we present the combined limits from the stacking algorithm. Finally, in Sect.~\ref{ch:systematic_uncertanties} we discuss the limitations of our results.

\subsection{Individual limits} \label{ch:individual_limits}

\begin{figure}[b]
    \centering
    \includegraphics[width=\hsize]{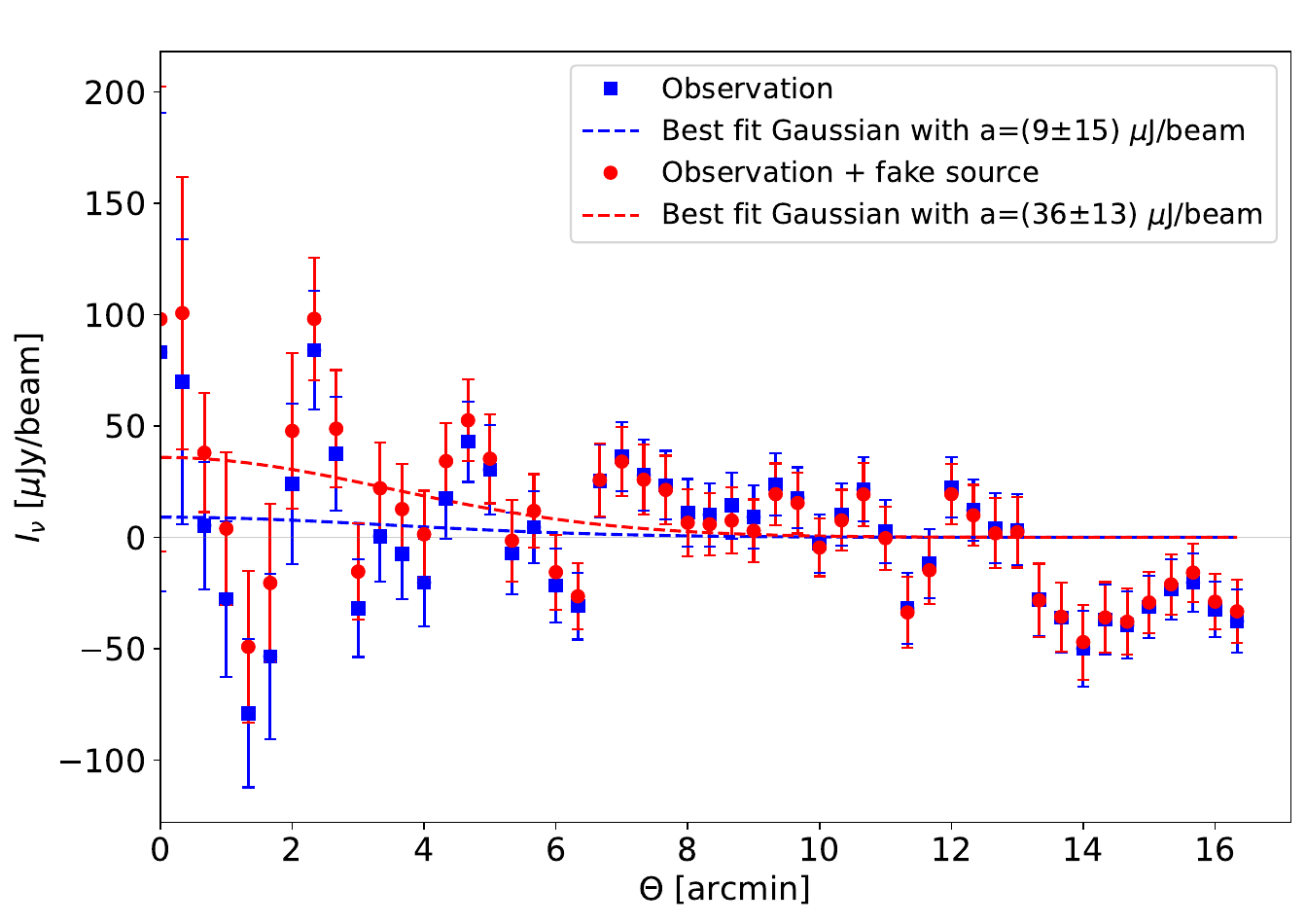}
    \caption{Radial intensity profiles for Canes Venatici I. Data points show the mean intensity in $20\arcsec$ wide annuli, with the error bars showing the standard deviation of the mean. Blue data points are for purely observational data, and red data points are for the same data with an additional $20~\rm mJy$ fake source. Dashed lines show the best-fitting Gaussians with a fixed FWHM of $r_\star$ equivalent to $8\farcm 2$.}
    \label{fig:CVNI_RP}
\end{figure}

Of the individual galaxies, we present first results of CVnI, which is already analyzed by \citet{vollmann20} using the same technique but with a slightly different implementation of other software. This galaxy serves as a benchmark to test our data processing algorithm. The radial intensity profiles with and without fake source are shown in Fig.~\ref{fig:CVNI_RP}. The amplitude of the Gaussian fitted to the observational data should be compatible with zero within the uncertainties to verify the non-detection of DM-related signals. Contrary, the amplitude of the Gaussian fitted to the data with the added fake source should be detected at 2$\sigma$ significance. This is indeed the case, as the best-fitting amplitude for the profile including the injected source is $\left(36\pm13\right) \muJybeam$ whereas without fake source it is $(9\pm 15)$~$\mu\rm Jy\,beam^{-1}$ at a FWHM of 8.2 arcmin. In Table \ref{tab:fake_source_fits}, we summarize the fitting results for each galaxy with the corresponding profiles presented in Appendix~\ref{a:individual_galaxies}.

 \begin{table}[t]
    \caption{Gaussian amplitude $a_{\text{Gauss, obs}}$ for purely observed data and amplitude $a_{\text{Gauss, inj}}$ for data with an additional injected fake source. $S_{\rm FS}$ is the fake source flux density, and FWHM is the width of the Gaussian, here assumed to be equivalent to $r_\star.$}
    \centering
    \begin{tabular}{lcccc}
     \hline \hline
        \noalign{\smallskip}
        dSph & $S_{\text{FS, $2\sigma$}}$ &  $a_{\text{Gauss, obs}}$& $a_{\text{Gauss, inj}}$ & FWHM\\
              & [mJy] & [$\muJybeam$] & [$\muJybeam$] & [arcmin]\\
        \hline
        \noalign{\smallskip}
        CVnI & 20 &$9\pm15$ & $36\pm13$ & 8.2\\
        UMaI & 35 & $-6\pm10$ & $24\pm11$ & 11.3\\
        UMaII & 35 & $-1.1\pm4.8$ & $12.9\pm4.8$ & 16.0\\
        UMi & 33 & $9.1\pm9.6$ & $28.9\pm9.0$ & 8.3 \\
        WilI & 5 & $-55\pm42$ & $17\pm39$ & 1.8 \\
        CVnII & 1 & $16\pm28$ & $50\pm29$ & 1.6 \\
        \hline
    \end{tabular}
    \label{tab:fake_source_fits}
\end{table}

Fluctuations that cannot be described by Gaussian statistics can affect the fit of galaxies with small half-light radii as the number of data points is small and any fluctuation may not average out. This is in particular the case for WilI ($r_\star = 110\arcsec$) and CVnII ($r_\star = 95\arcsec$). For WilI, a negative Gaussian amplitude for the fit to the purely observational data is found. This is the only galaxy with a "signal," but because it is negative, it can be ruled out as DM-related. For CVnII, the Gaussian amplitude is compatible with zero, albeit with a large uncertainty. We tried to mitigate the limitation due to the small number of data points by increasing the region in which radial profiles were measured to  $3r_\star$ and $4r_\star$ for WilI and CVnII, respectively.

\begin{figure}[t]
    \centering
    \includegraphics[width=\hsize]{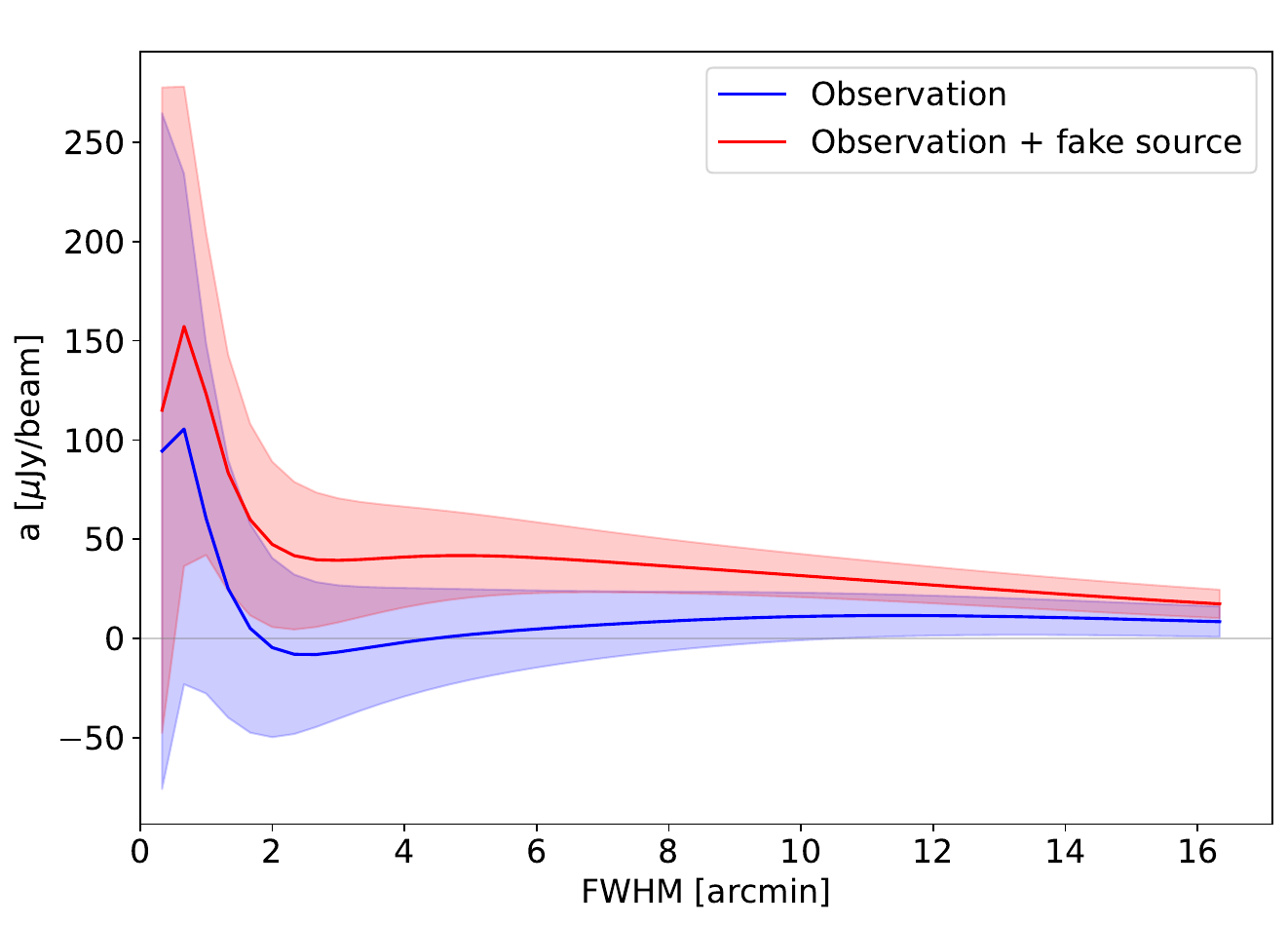}
    \caption{Best-fitting Gaussian amplitudes for the radial intensity profiles of Canes Venatici I with Gaussians of varying FWHMs. The solid blue line is for purely observational data, and the solid red line the same data with an additional $20$~mJy fake source. Shaded areas indicate $1\sigma$ confidence intervals.}
    \label{fig:CVNI_FWHM}
\end{figure}

Since the size of the DM halo is uncertain, we varied the FWHM of the Gaussian fit to the radial intensity profiles. Here, we wanted to investigate the possible systematic uncertainty of the assumption that the FWHM of the signal produced by the DM halo is equal to the stellar radius, $r_\star$, which itself has an uncertainty of around 10 to 15\% \citep{geringer15}. For CVnI, the corresponding results are shown in Fig.~\ref{fig:CVNI_FWHM}. The data are in agreement with zero within the 1$\sigma$ confidence intervals for almost the entire range of FWHM. On the other hand, the $2\sigma$ detection of the injected source over a wide range of FWHM is also evident. Only at small FWHM, statistical fluctuations start to suppress the significance of the detection. Signals on that scale are most likely due to fluctuations in the map and not related to DM.

The next step was to calculate the diffusion and energy-loss timescales. This identified the diffusion regime that then lead to the set of equations needed to estimate limits on the annihilation cross section. For the INT scenario, the diffusion timescale for CVnI is $\approx$30~Myr, whereas the energy-loss timescale is $\approx$110 Myr. Since both timescales are on the same order of magnitude, we used diffusion regime $B$ with Eq.~\eqref{eq:spectral_part_B} to calculate the limits on the WIMP annihilation cross section. We summarize the diffusion and energy-loss timescales together with the resulting diffusion regimes for the three model scenarios for our six dSph galaxies in Table~\ref{tab:time_scales}. We note that both timescales depend on the CR energy. We used a benchmark-energy of $E = 10~\GeV$; a different CR energy may change the choice of diffusion regimes and hence slightly affect the limits. 

\begin{table}[t]
    \caption{Timescales of CR diffusion and loss for every model scenario with a benchmark CR energy of 10 GeV. The resulting diffusion regime is also noted.}
    \centering
    \begin{tabular}{lcccc}
     \hline \hline
        \noalign{\smallskip}
        dSph & Model scenario & $\tau_{\text{diff}}$ & $\tau_{\text{loss}}$ & Regime\\
               & & [Myr] & [Myr] & \\
        \hline
        \noalign{\smallskip}
        CVnI & OPT & 288 & 11.4 & $A$\\
         & INT & 28.8 & 113 & $B$\\
         & PES & 0.288 & 124 & $C$\\
        \hline
        \noalign{\smallskip}
        UMaI & OPT & 92.1 & 11.4 & $A$ \\
         & INT & 9.21 & 113 & $C$\\
         & PES & 0.0921 & 124 & $C$\\
        \hline
        \noalign{\smallskip}
        UMaII & OPT & 20.1 & 11.4 & $B$ \\
        & INT & 2.01 & 113 & $C$\\
        & PES & 0.0201 & 124 & $C$\\
        \hline
        \noalign{\smallskip}
        UMi & OPT & 29.7 & 11.4 & $B$ \\
        & INT & 2.97 & 113 & $C$\\
        & PES & 0.0297 & 124 & $C$\\
        \hline
        \noalign{\smallskip}
        WilI & OPT & 0.399 & 11.4 & $C$ \\
        & INT & 0.0399 & 113 & $C$ \\
        & PES & 0.000399 & 124 & $C$ \\
        \hline
        \noalign{\smallskip}
        CVnII & OPT & 4.96 & 11.4 & $B$ \\
        & INT & 0.496 & 113 & $C$\\
        & PES & 0.00496 & 124 & $C$\\
        \hline
    \end{tabular}
    \label{tab:time_scales}
\end{table}

\begin{figure*}[h!]
    \centering
     \begin{subfigure}[t]{0.02\textwidth}
        \textbf{(a)}    
    \end{subfigure}
    \begin{subfigure}[t]{0.47\textwidth}
        \centering
        \includegraphics[width=\linewidth,valign=t]{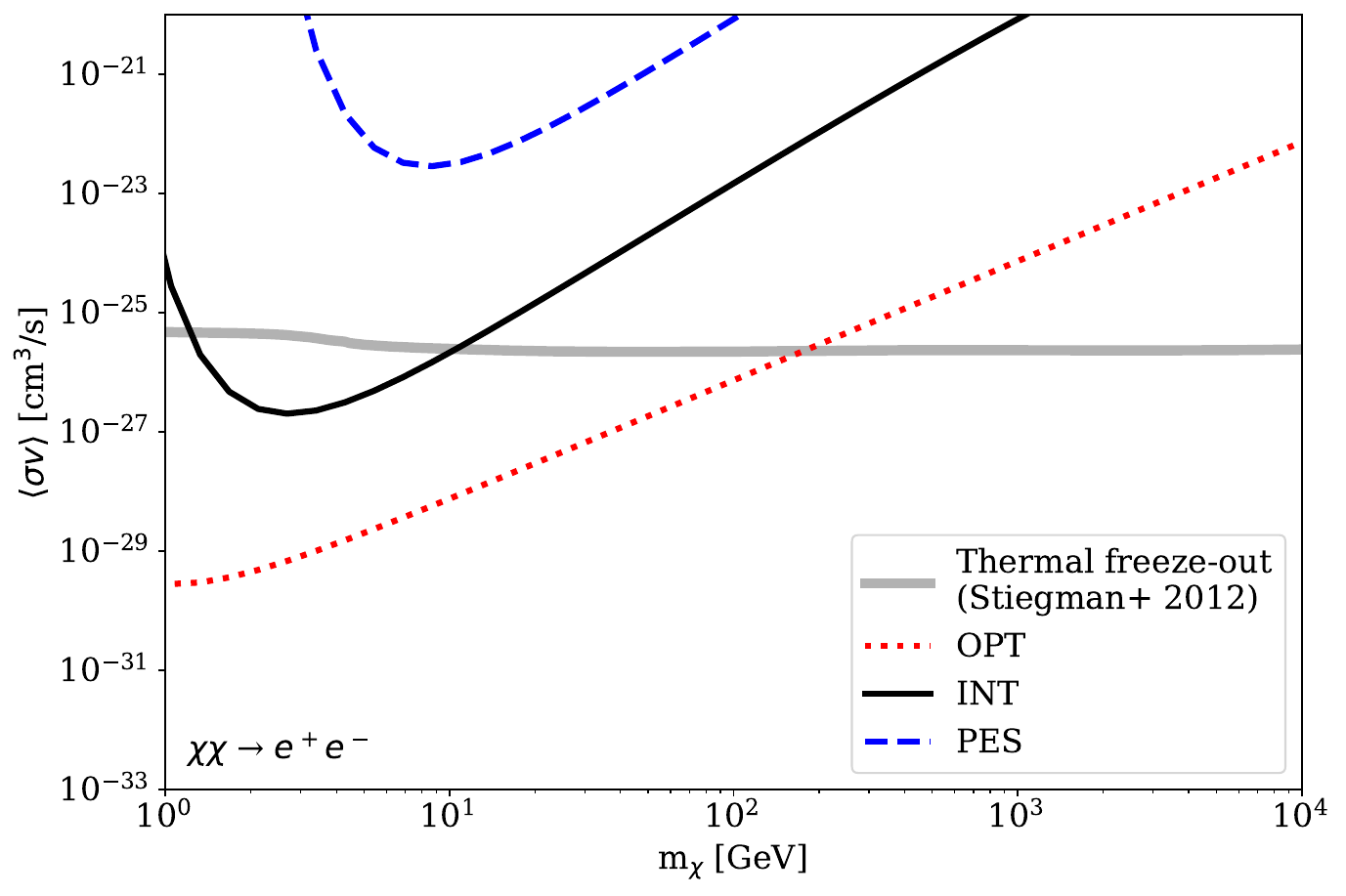}
    \end{subfigure}%
    \begin{subfigure}[t]{0.02\textwidth}
        \textbf{(b)}    
    \end{subfigure}
    \begin{subfigure}[t]{0.47\textwidth}
        \centering
        \includegraphics[width=\linewidth,valign=t]{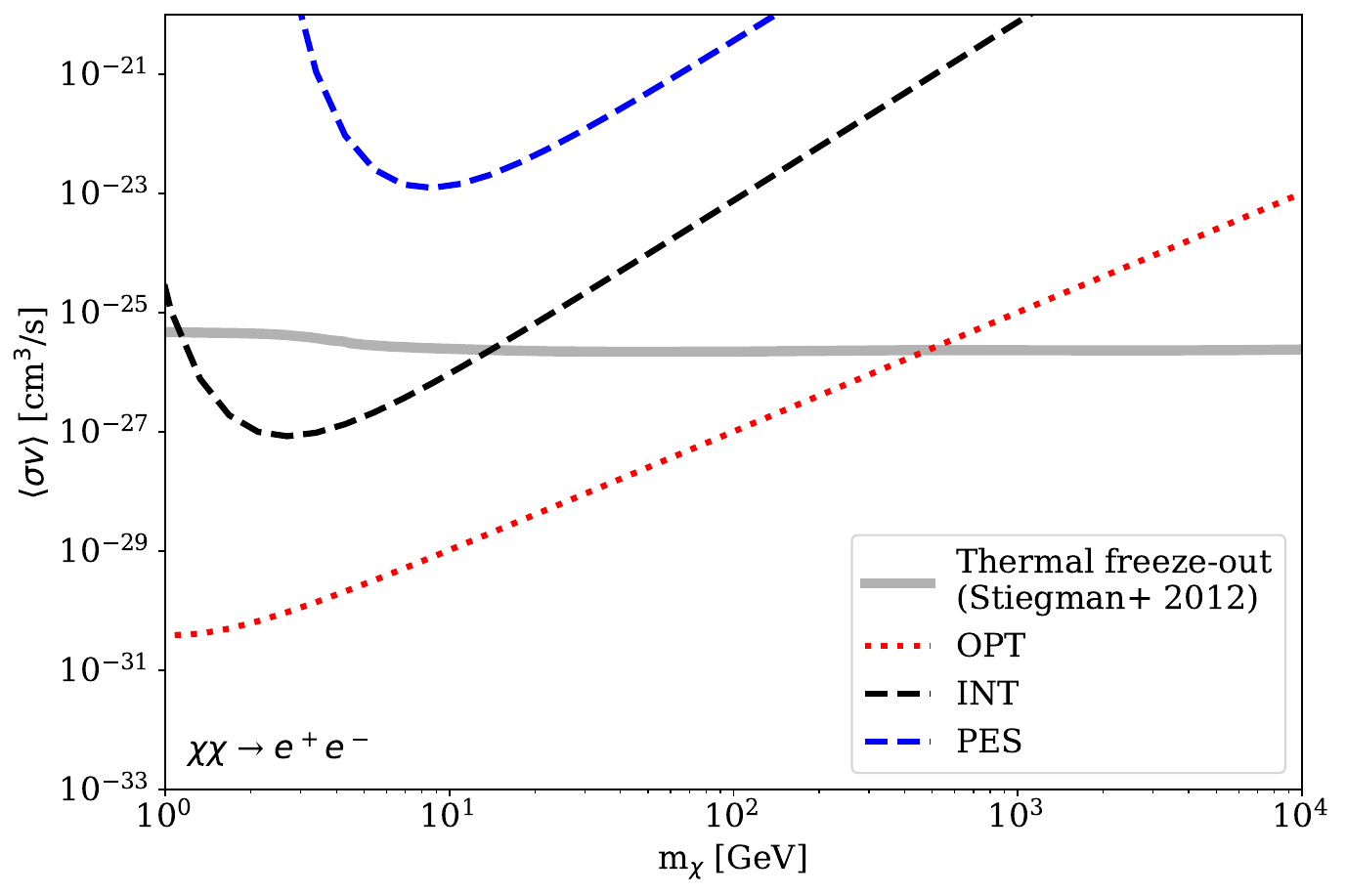}
    \end{subfigure}
    \\
    \begin{subfigure}[t]{0.02\textwidth}
        \textbf{(c)}    
    \end{subfigure}
    \begin{subfigure}[t]{0.47\textwidth}
        \centering
        \includegraphics[width=\linewidth,valign=t]{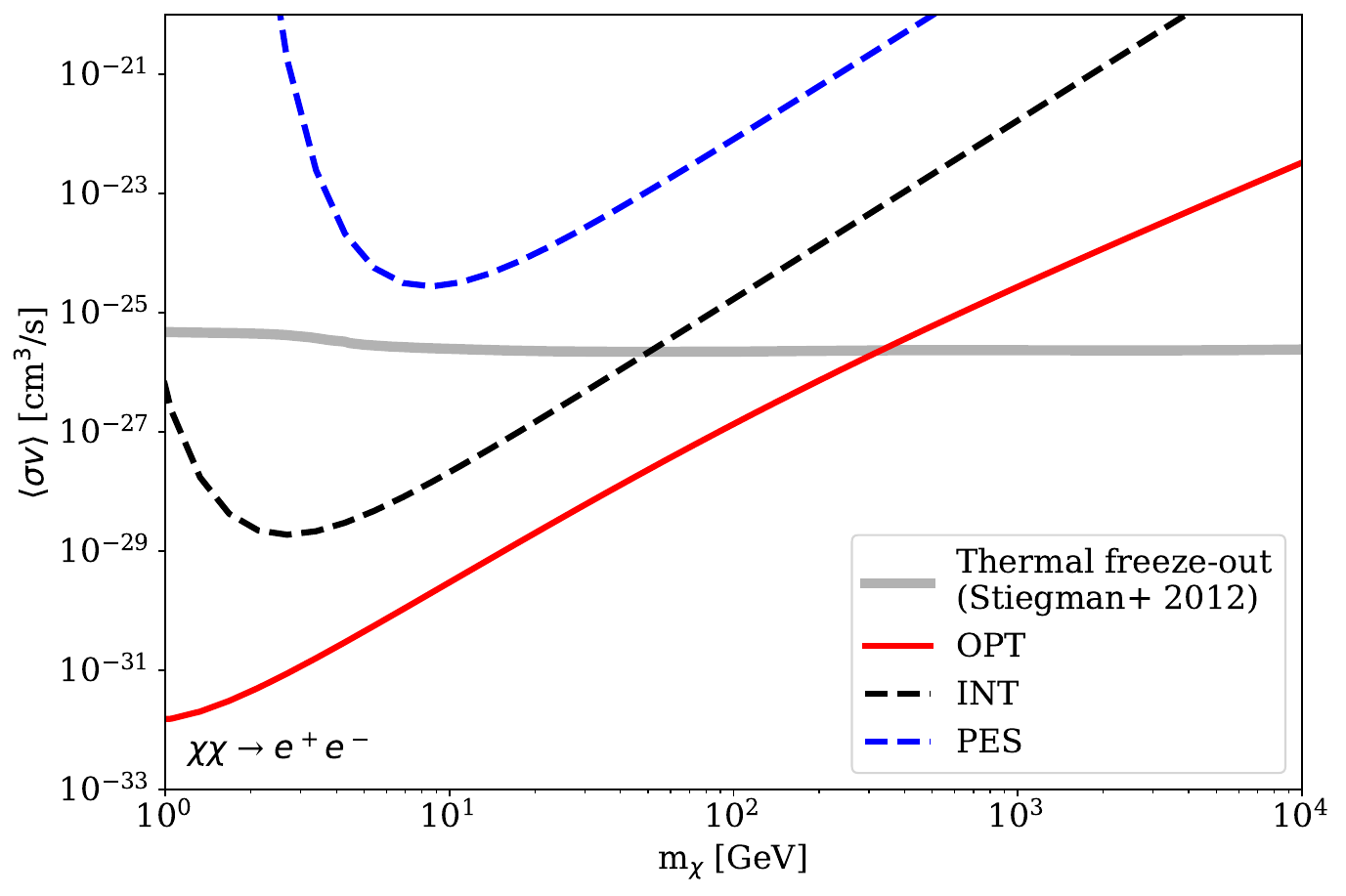}
    \end{subfigure}%
     \begin{subfigure}[t]{0.02\textwidth}
        \textbf{(d)}    
    \end{subfigure}
    \begin{subfigure}[t]{0.47\textwidth}
        \centering
        \includegraphics[width=\linewidth,valign=t]{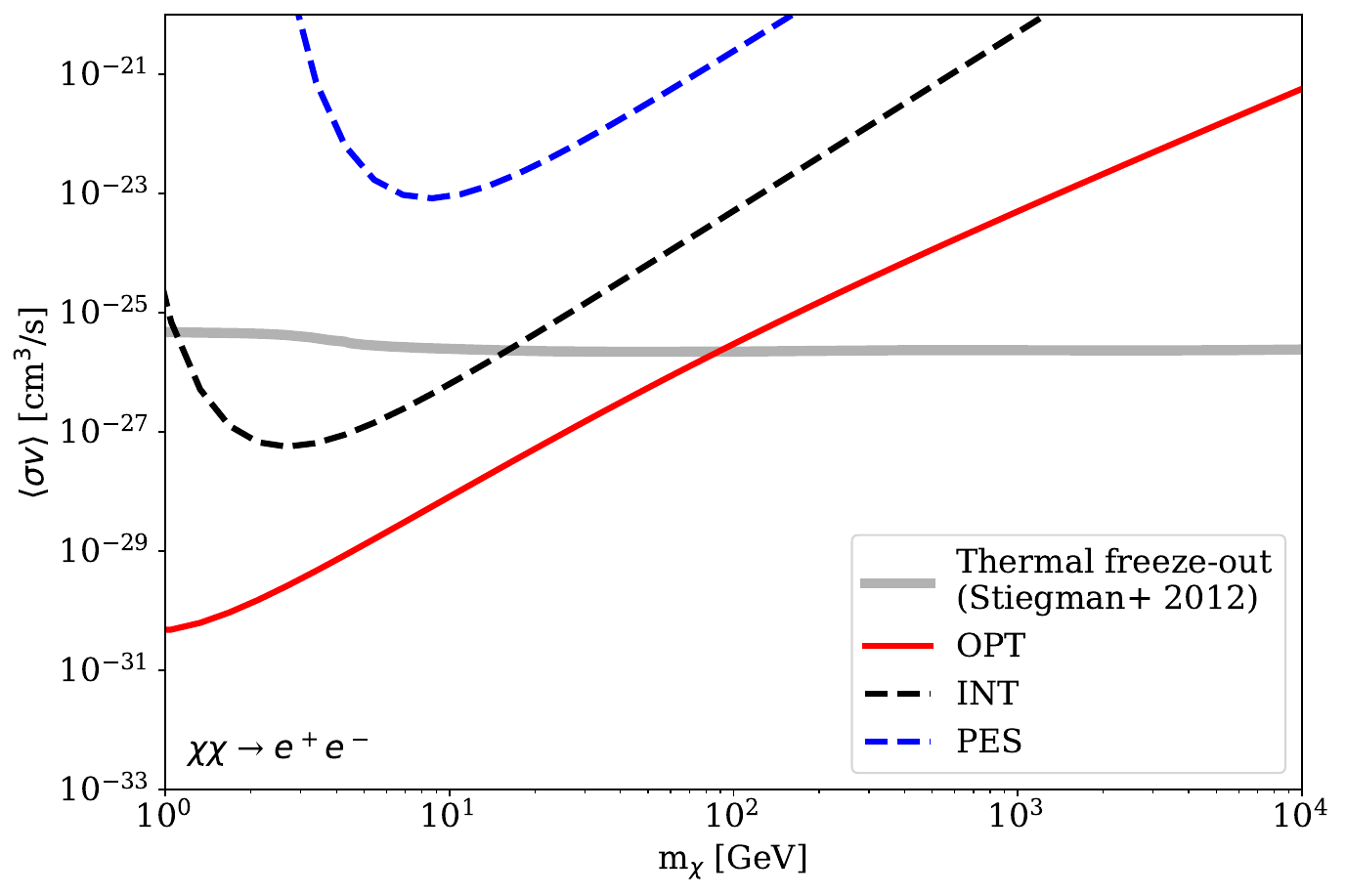}
    \end{subfigure}
    \\
    \begin{subfigure}[t]{0.02\textwidth}
        \textbf{(e)}    
    \end{subfigure}
    \begin{subfigure}[t]{0.47\textwidth}
        \centering
        \includegraphics[width=\linewidth,valign=t]{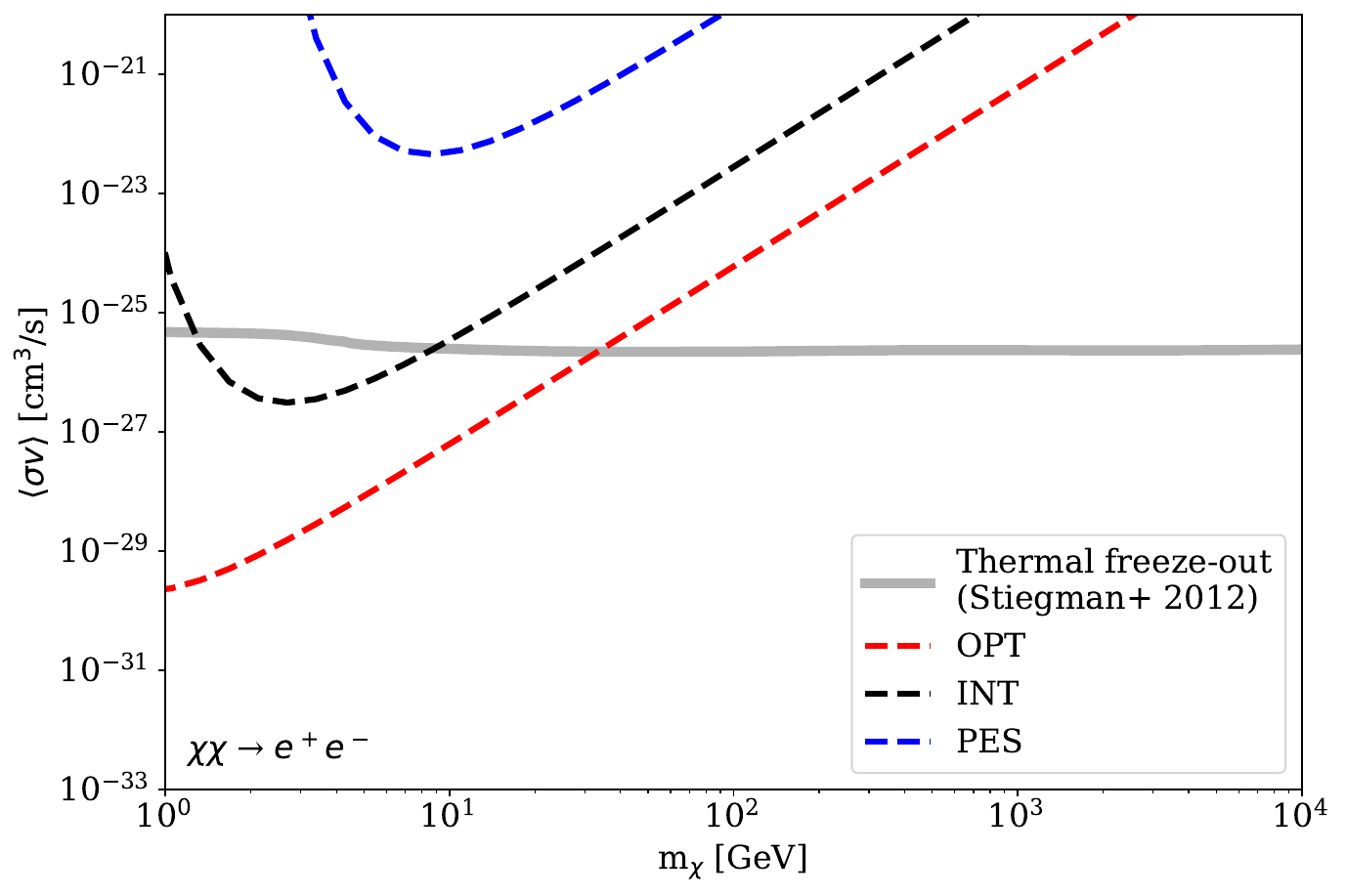}
    \end{subfigure}%
    \begin{subfigure}[t]{0.02\textwidth}
        \textbf{(f)}    
    \end{subfigure}
    \begin{subfigure}[t]{0.47\textwidth}
        \centering
        \includegraphics[width=\linewidth,valign=t]{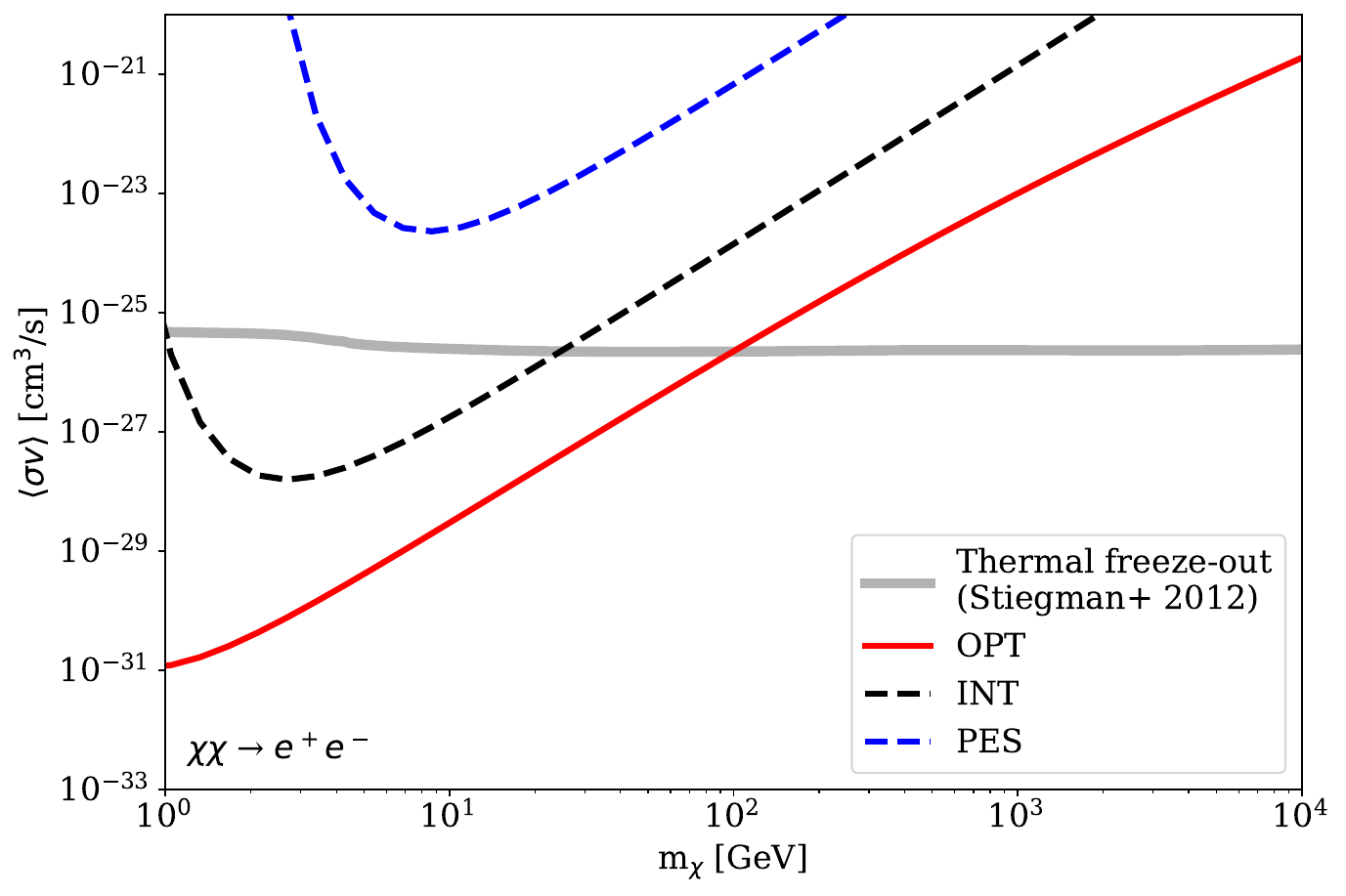}
    \end{subfigure}
    \caption{Individual upper limits on the WIMP annihilation cross section for the OPT, INT, and PES model scenarios. Line colors represent the assumed model scenario: red for OPT, black for INT, and blue for PES. Line styles represent the assumed diffusion regime: dashed lines for regime $A$, solid lines for regime $B,$ and dotted lines for regime $C$. Panel (a) shows CVnI, (b) UMaI, (c) UMaII, (d) UMi, (e) WilI, and (f) CVnII. The gray line represents the lower limit from the thermal freeze-out \citep{stiegman12}. }
    \label{fig:sigmav_plots}
\end{figure*}

In Fig.~\ref{fig:sigmav_plots} we present the upper limits to the WIMP annihilation cross section from each individual galaxy. For comparison, we additionally plot the lower limit on the annihilation cross section calculated from the thermal WIMP freeze-out mechanism by \citet{stiegman12}. 

\subsection{Stacked limits} \label{ch:stacked_limits}

\begin{figure*}[t]
    \centering
     \begin{subfigure}[t]{0.02\textwidth}
        \textbf{(a)}    
    \end{subfigure}
    \begin{subfigure}[t]{0.47\textwidth}
        \centering
        \includegraphics[width=\linewidth,valign=t]{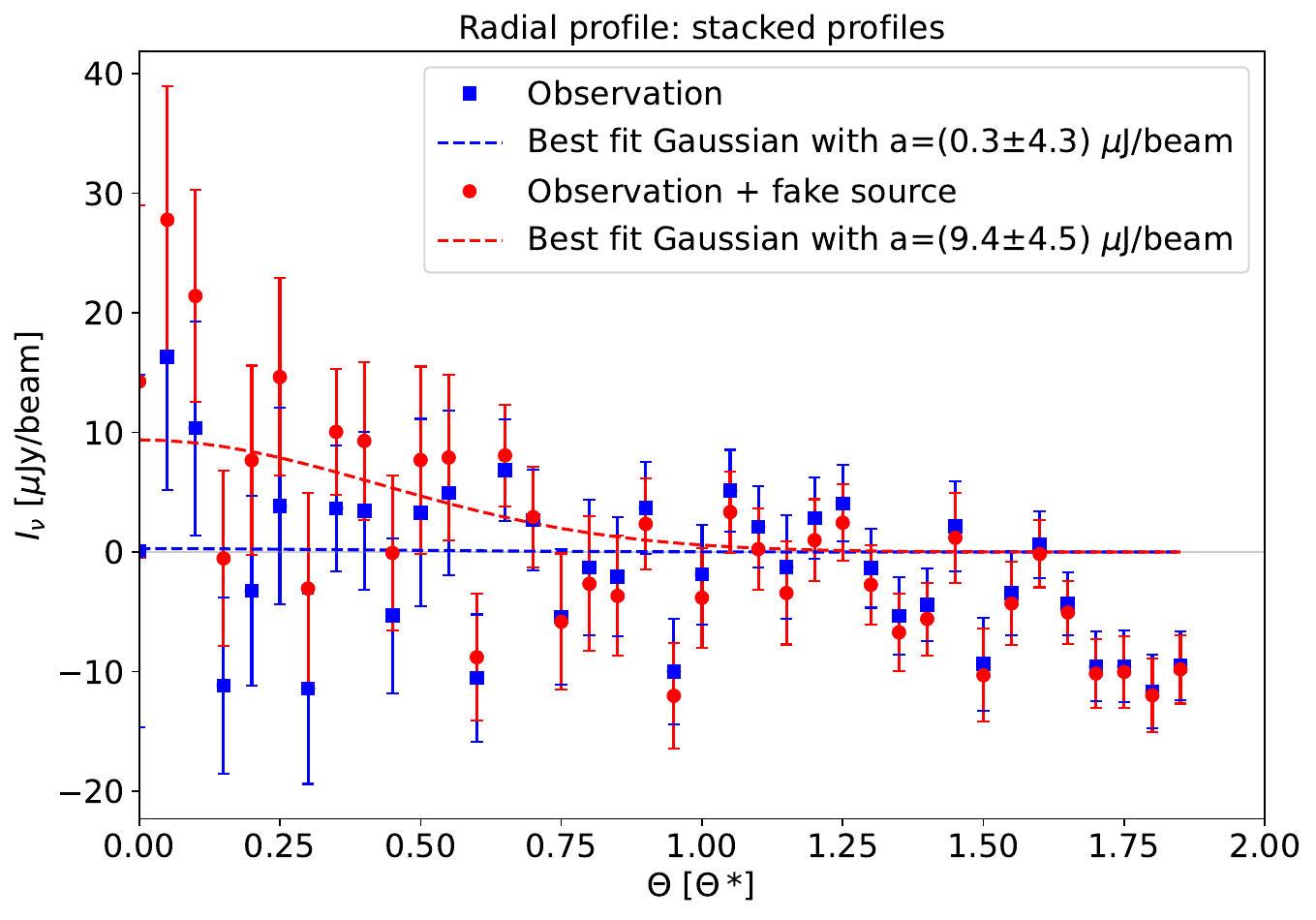}
    \end{subfigure}%
     \begin{subfigure}[t]{0.02\textwidth}
        \textbf{(b)}    
    \end{subfigure}
    \begin{subfigure}[t]{0.47\textwidth}
        \centering
        \includegraphics[width=\linewidth,valign=t]{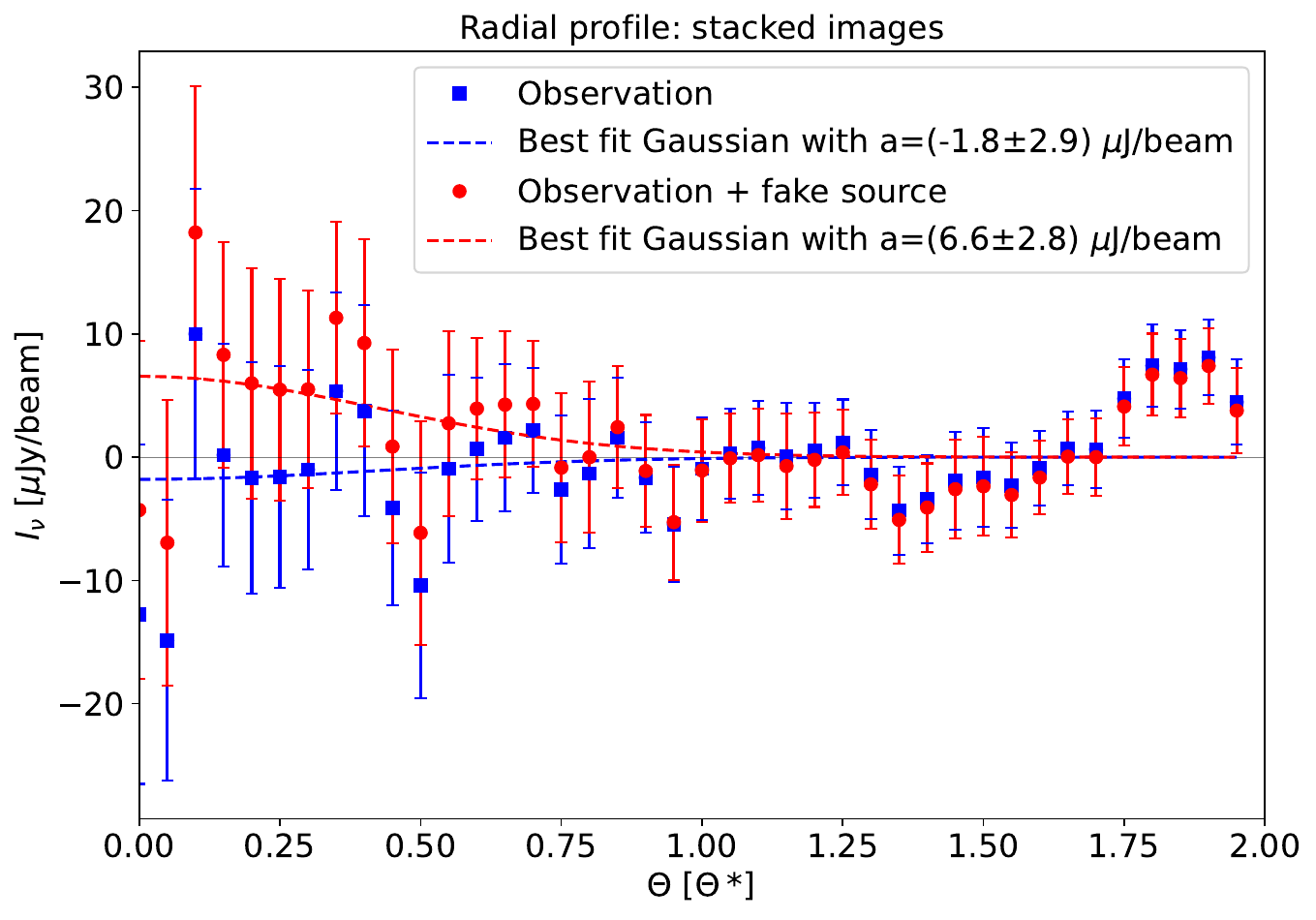}
    \end{subfigure}
    \caption{Stacked radial intensity profiles. Panel (a) shows profile stacking and panel (b) image stacking. Data points show the mean intensity in adjacent annuli, with the error bars showing the standard deviation of the mean. Blue data points are for purely observational data, and red data points are for the same data with an additional fake source. Dashed lines show the best-fitting Gaussians with a fixed FWHM of $r_\star$. The radius is expressed as an apparent angle, $\theta$, scaled to the apparent size of the stellar radius, $\theta_\star=r_\star/d$.}
    \label{fig:stacked_RP_appendix}
\end{figure*}

\begin{figure*}[t]
    \centering
     \begin{subfigure}[t]{0.02\textwidth}
        \textbf{(a)}    
    \end{subfigure}
    \begin{subfigure}[t]{0.47\textwidth}
        \centering
        \includegraphics[width=\linewidth,valign=t]{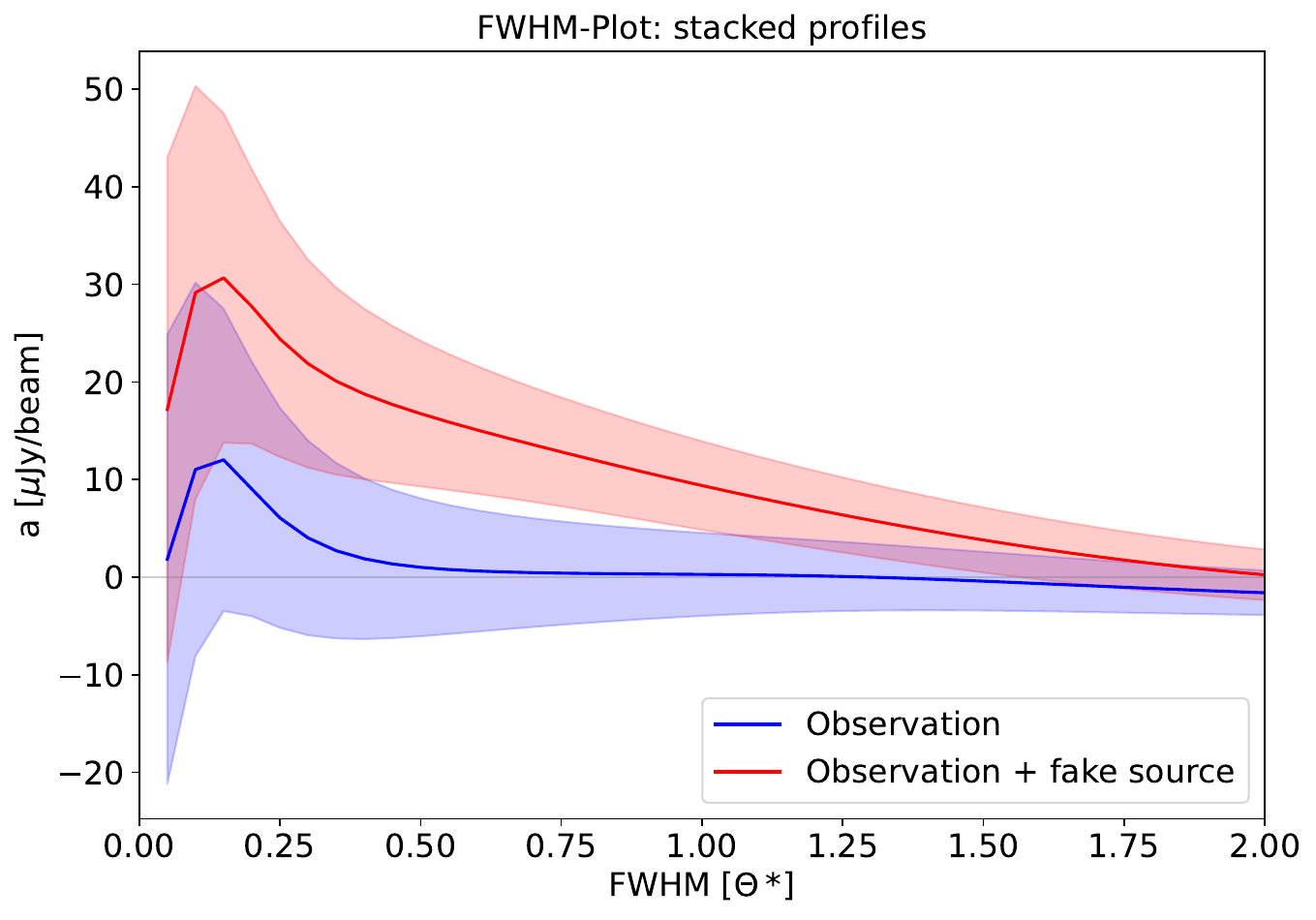}
    \end{subfigure}%
     \begin{subfigure}[t]{0.02\textwidth}
        \textbf{(b)}    
    \end{subfigure}
    \begin{subfigure}[t]{0.47\textwidth}
        \centering
        \includegraphics[width=\linewidth,valign=t]{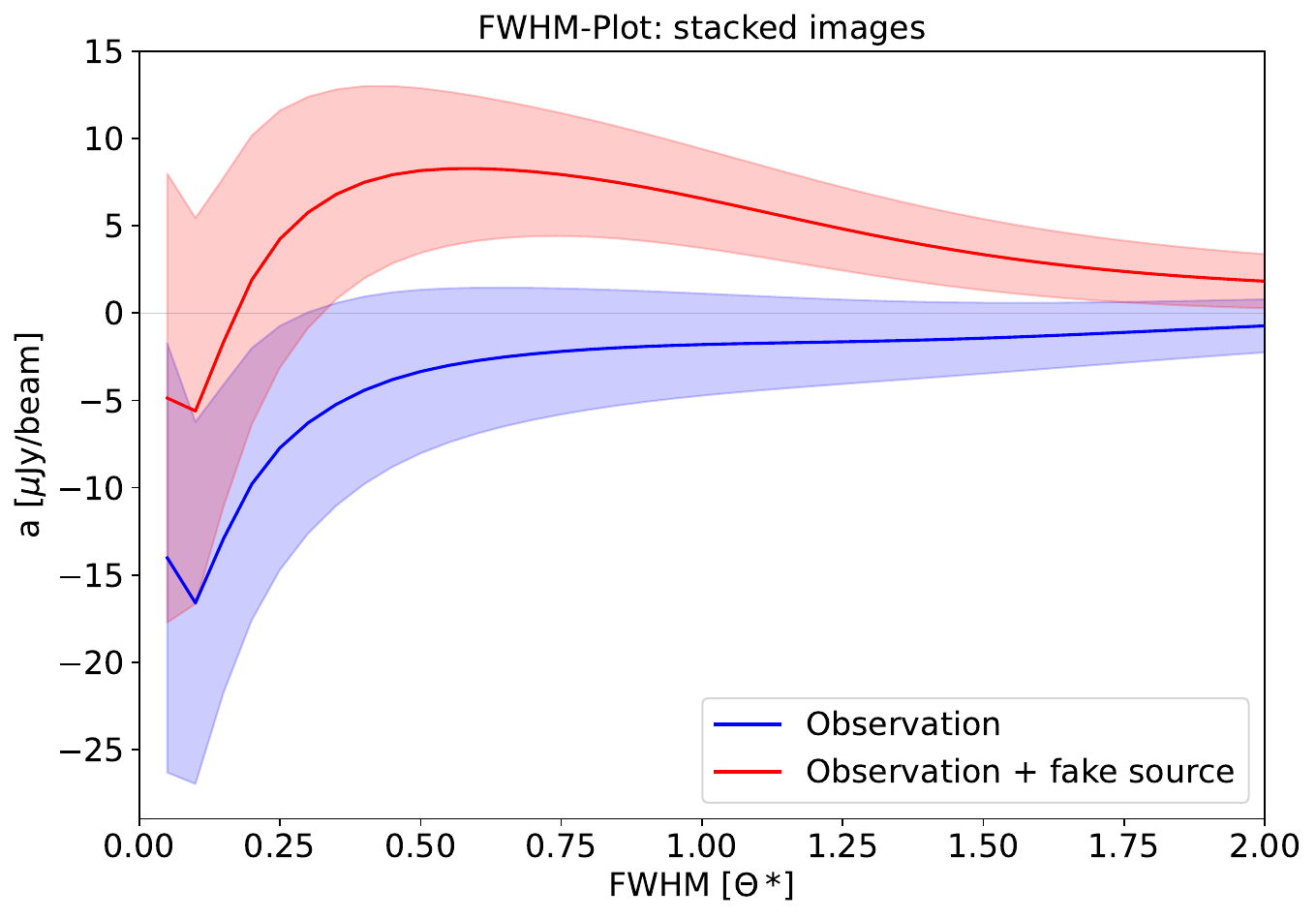}
    \end{subfigure}
    \caption{Best-fitting Gaussian amplitudes for the stacked radial intensity profiles. Panel (a) shows profile stacking and panel (b) image stacking. Solid blue lines show the purely observational data, whereas solid red lines show the same data with an additional fake source. Shaded areas indicate $1\sigma$ uncertainties. The FWHM of the Gaussian is expressed as the apparent size of the stellar radius, $\theta_\star=r_\star/d$.}
    \label{fig:stacked_FWHM_appendix}
\end{figure*}

The stacked radial intensity profiles both for profile and image stacking are shown in Fig. \ref{fig:stacked_RP_appendix}. The best-fitting Gaussian amplitudes are $a_{\text{obs, profiles}}=(0.3\pm4.3)\ \muJybeam$ and $a_{\text{obs, images}}=(-1.8\pm2.9)\ \muJybeam$ for the purely observed data using profile and image stacking, respectively. In both stacking strategies, the amplitudes are consistent with zero so the stacking does not reveal any additional signal. We varied the FWHM of fitted Gaussians as before for the individual galaxies. Again, the observed amplitudes are consistent with zero as shown in Fig. \ref{fig:stacked_FWHM_appendix}.

The flux density of the injected fake sources from the individual galaxies was divided by a factor to determine how much fainter is the signal that can be detected by stacking. The factor was chosen to achieve a 2$\sigma$ detection, it equals 1.8 for profile stacking and 2.8 for image stacking. We again fitted Gaussians to the resulting stacked radial profile and the Gaussian amplitudes are $a_{\text{inj, profiles}}=(9.4\pm4.5)\ \muJybeam$ for profile stacking and $a_{\text{inj, images}}=(6.6\pm2.8)\ \muJybeam$ for image stacking, showing that we have detected the fake source at $2\sigma$ confidence.

\begin{figure}
    \centering
    \includegraphics[width=\hsize]{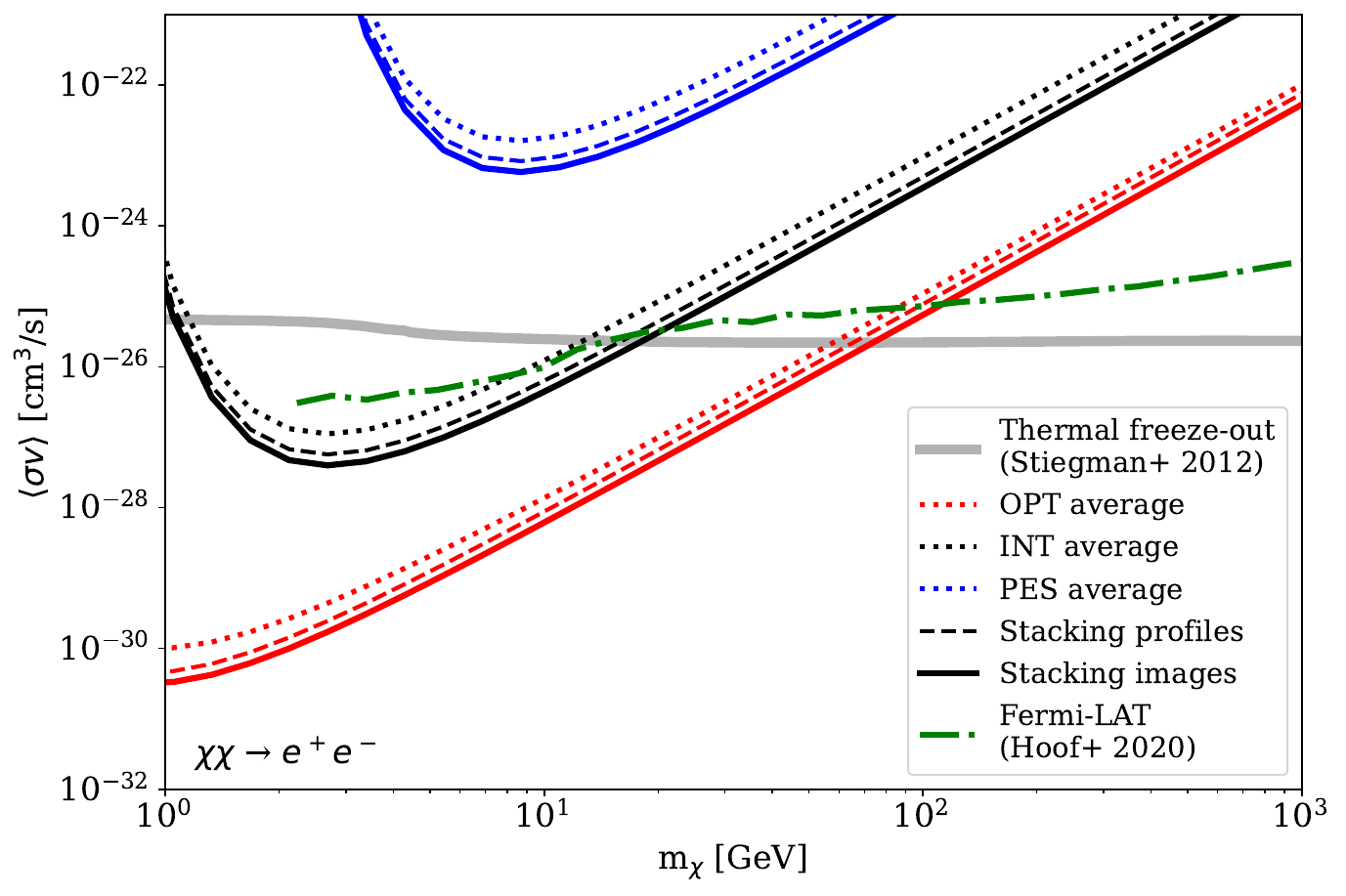}
    \caption{Upper limits on the WIMP annihilation cross section from stacking. Dotted lines are the averaged individual limits, dashed lines are limits obtained by stacking radial profiles, and solid lines are limits obtained by stacking galaxy images. Red, black, and blue correspond to the OPT, INT, and PES model scenarios, respectively. The thermal freeze-out cross section \citep[][gray]{stiegman12} and upper limits from {\it Fermi}--LAT $\gamma$-ray observations \citep[][green]{hoof20} are shown for comparison.}
    \label{fig:stacked}
\end{figure}

Upper limits for the WIMP annihilation cross section were calculated from the best-fitting amplitudes of the radial profiles obtained with either method. A unique value of amplitude $a$ was used and the other terms in Eq.~\eqref{eq:cross-section} were given as an average value for all galaxies. This average was calculated using the same approximation regime as for the individual galaxies (Table~\ref{tab:time_scales}). The resulting upper limits for the cross section from both approaches are shown in Fig.~\ref{fig:stacked}, together with the average of the limits obtained for the individual galaxies.

\subsection{Systematic uncertainties} \label{ch:systematic_uncertanties}

There are several sources of inaccuracies in the obtained limits. The most significant are the assumptions on the values of the average magnetic field strength and diffusion coefficient. For this reason we used our three different model scenarios in order to illustrate the influence of these parameters. The differences between the cross-section limits in each scenario are indeed two to three orders of magnitude showing the large uncertainty resulting from the inaccuracy of these input parameters.

For our INT scenario we assumed that the diffusion coefficient is $10^{27}$ cm$^2$s$^{-1}$ as measured both in dwarf irregular galaxies \citep{murphy12} and in the Milky Way \citep{korsmeier16}. However, it is also possible that the true value may be closer to $10^{28}$ cm$^2$s$^{-1}$ as found in late-type spiral galaxies \citep{heesen19} or even $10^{29}$ cm$^2$s$^{-1}$ such as for the halos in edge-on spiral galaxies. On the other hand, the true value can also go down to the value of $10^{26}$ cm$^2$s$^{-1}$ measured in dwarf irregular galaxies \citep{heesen18}. The magnetic field is much more uncertain but values range between 0.1 $\mu$G and 10 $\mu$G assuming that the magnetic energy density is in equipartition with the CR energy density within two orders of magnitude. We note that this is a only a heuristic argument though as there are no stringent physical reasons why there should be equipartition. Generally, the diffusion coefficient depends on the magnetic field \citep{sigl17}, but to simplify our model we treated them independently and this is an additional source of uncertainty. 

A further uncertainty is introduced by the adopted NFW profile parameters $r_s$ and $\rho_s$. While including full posterior probability-distribution functions (PDFs) of existing Bayesian fits in the literature \citep[e.g.,][]{ando20} would be more appropriate, it would also be impractical. This is because of the overwhelming magnetic-field uncertainties, which would not be reduced by considering such PDFs. Therefore, we contend ourselves by using the best-fit values for $r_s$ and $\rho_s$ as derived in \citep{geringer15}, and direct our attention to the CR propagation parameters (e.g., magnetic field) when discussing uncertainties.

Another source of uncertainty comes from the approximations made in the spectral function where we chose one of the diffusion regimes ($A$, $B$, or $C$) and used the appropriate equation (Eq.~\ref{eq:spectral_part_A}, \ref{eq:spectral_part_B}, or \ref{eq:spectral_part_C}). The regime was chosen by comparing the diffusion and energy-loss timescales (Table \ref{tab:time_scales}). The diffusion timescale depends on the diffusion coefficient, $D_0$, and the energy-loss timescale depends on the average magnetic field, $B$. For this reason, regimes were chosen independently for each model scenario. An inappropriate choice of the approximation regime for certain ratio of timescales affects the cross-section limits much less than the previously mentioned uncertainty from the model parameters ($D_0$, $B$). There is ongoing work to solve the transport equations regardless of the ratio of timescales \citep{vollmann22}, which would completely eliminate such uncertainties.

There is some evidence for tidal disruption or nonequilibrium kinematics in WilI \citep{ibata97, willman11, geringer15}, which could cause the NFW profile parameters to be biased to higher masses. Since that bias might artificially improve our stacked limits, we repeated the stacked analysis without WilI but with the same injected source intensity for other galaxies (see Appendix~\ref{a:noWLMI}). We find that the resulting limits without WilI are stronger compared to the limits including all the galaxies. In the end, we decided to keep the weaker limits derived from all galaxies as the more conservative estimate.

We also mention other sources of uncertainty that are small compared to the model scenario uncertainty. First, we assumed that the density distribution in dSph galaxies is described by a NFW density profile. This is just one of the possibilities and alternative profiles \citep[e.g.,][]{einasto89,burkert95} could also be used \citep{vollmann21}. Second, we assumed a power-law dependence of the diffusion coefficient on the CR energy (Eq.~\ref{eq:diffusion_coeff}) and Kolmogorov turbulence. If we are able to better constrain the magnetic field and diffusion coefficient in dSph galaxies, the mentioned uncertainties would become more important.

\section{Discussion and conclusions}
\label{s:disccusion_and_conclusions}

\begin{figure}
    \centering
    \includegraphics[width=\hsize]{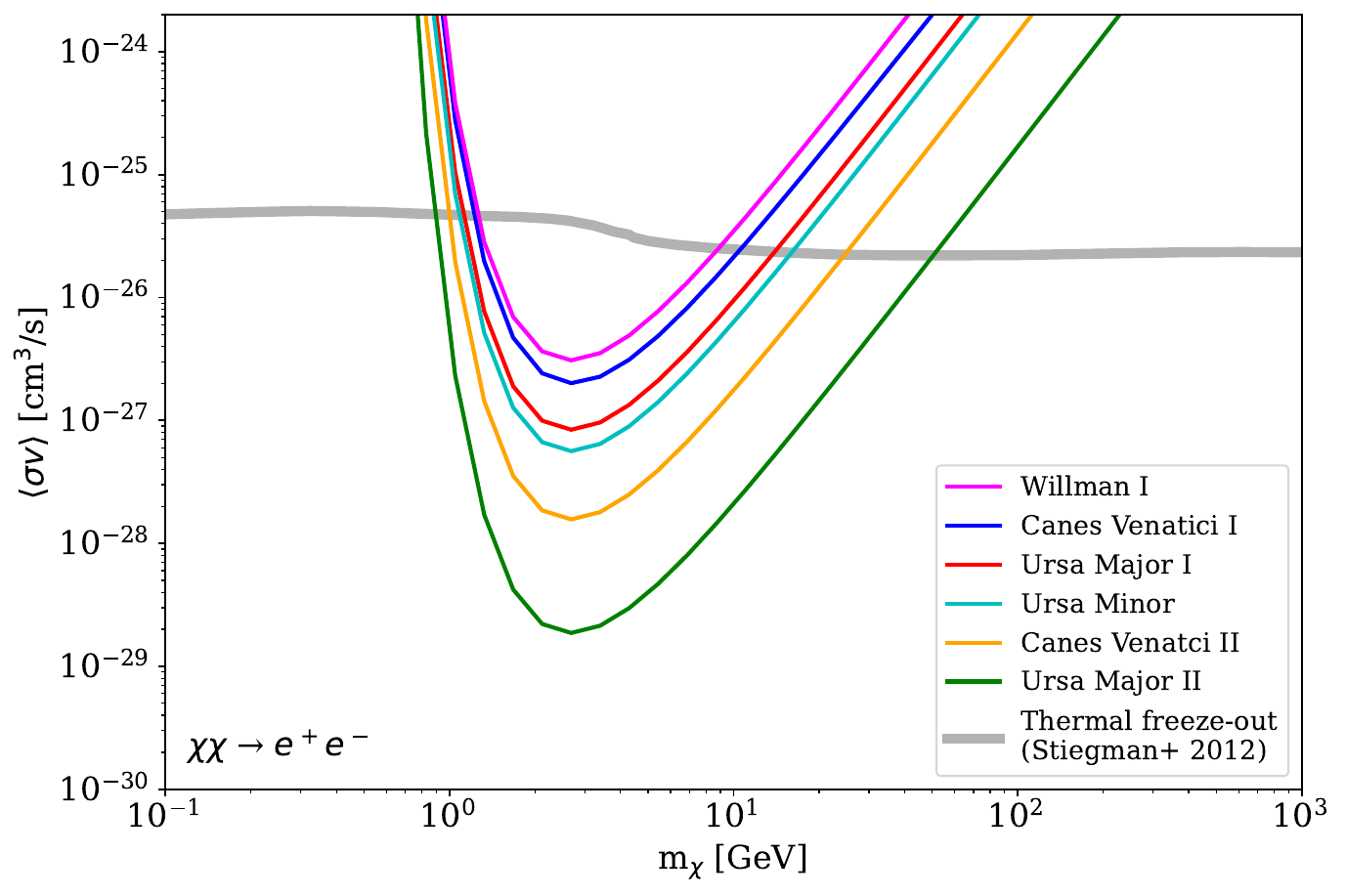}
    \caption{Individual upper limits on the WIMP annihilation cross section for the dSph galaxies in our sample for the INT model scenario. }
    \label{fig:gal_comparison}
\end{figure}

We first compared the limits for the INT model scenario from individual galaxies in Fig.~\ref{fig:gal_comparison}. The individual limits within our sample vary due to the different noise levels of the LoTSS maps and the different intrinsic properties, such as the DM density profile and distance. 
The best constraints on the annihilation cross section are derived from UMaII. This is also true for $\gamma$-ray WIMP searches since this galaxy has the highest $J$ factor \citep{hoof20}. The next best constraints are derived from CVnII. This is counterintuitive since UMaII has the largest apparent size ($r_\star \approx 16\arcmin$), whereas CVnII is the smallest ($r_\star \approx 1\farcm 6$). But reviewing the halo factor defined in Eq.~\eqref{eq:halo_part}, it becomes evident that the apparent size is not the most important parameter; the most important are instead the characteristic density, $\rho_s$, the scale length, $r_s$, and the half-light radius, $r_\star$. According to Eq.~\eqref{eq:halo_part}, the combination of these parameters is high for UMaII and CVnII, which explains the strong limits for these two galaxies. The same explanation justifies why the limits set by WilI and CVnI are less stringent than average: WilI has the lowest scale length within our set of galaxies\footnote{The spectral factor $X(\nu)$ for WilI is two orders of magnitudes smaller than average, which justifies the moderate limit for this galaxy as well.}, and CVnI has the smallest characteristic density. Since the halo factor depends on the square of both values, this factor is smaller than average, and hence the limits are less stringent. 

We improved the limits by stacking the data for different galaxies. To test the improvement, we compared the stacked limits to the average value of all individual limits (Fig.~\ref{fig:stacked}). The stacking of profiles lowered the limits by a factor of approximately two and the stacking of images by three. This is expected because with stacking we are effectively extending the observing time. 
The image stacking yields better results than the stacking of profiles, but controlling the uncertainties is more difficult. We only compared the two specific stacking strategies, but in general there are many alternative ways of combining the results of different galaxies. For example, in a future study a more statistically rigorous approach could be to treat galaxies independently with their own nuisance parameters while the DM model parameters are fitted simultaneously \citep[see, e.g.,][]{hoof20}.

The comparison of our stacking limits with other attempts to constrain the WIMP annihilation cross section, such as the thermal freeze-out cross section by \citet{stiegman12} or dSph galaxy observations by {\it Fermi}--LAT \citep{hoof20}, shows the competitiveness of our results. Our cross-section limits from the INT scenario already exclude thermal WIMPs with masses below 20~GeV. In the OPT scenario we can even exclude thermal WIMPs with masses below 70~GeV. While the PES limits do not exclude thermal WIMPs in any mass range, they still prove the validity of our concept \citep{vollmann20, regis14}.

As is customary, we assumed a smooth DM distribution. However, there has been recent work stating that DM halos contain prompt cusps \citep{delos22} that would boost the annihilation signal, which, in turn, would lower our limits by up to two orders of magnitude. The same effect would apply to limits inferred from $\gamma$-ray excess in the Milky Way and other galaxies.

To date, the results from $\gamma$-ray observations of dSph galaxies presented by \cite{hoof20} are among the strongest limits on the WIMP annihilation cross section. In the mass range below 20~GeV, our INT limits are stronger than those from {\it Fermi}--LAT. Our limits are valid under the assumption of the average magnetic field and diffusion coefficient in dSph galaxies and therefore have a large uncertainty. However, our results show that radio observations of dSph galaxies can potentially constrain the WIMP annihilation cross section if one accepts this premise. This method is especially powerful for WIMP masses below 10~GeV. 

Considering the observational resources needed for both attempts, our method is much more efficient. We only used LoTSS-DR2 survey data \citep[][about 50~h in total for the six dSph galaxies]{shimwell22} and have not performed targeted observations. For comparison, \cite{ackermann15} used six years of {\it Fermi}--LAT data. Compared to these observations, we achieve better limits at lower masses in the OPT scenario and comparable limits in the INT scenario. Hence, the advantages of our study are the sensitivity in the low-mass regime and the efficiency in terms of observation time. The biggest drawbacks of using radio continuum observations are of course the uncertainties related to the strength of the magnetic field and the value of the diffusion coefficient. The field strength of dSph galaxies could be measured from a grid of Faraday rotation measures of polarized background sources with a sensitive radio telescope, such as the Square Kilometre Array \citep[SKA;][]{johnston15}.

In addition to the HBA used for LoTSS, LOFAR low band antenna observations \citep{degasperin21} would improve the limits on the lower mass end because the critical frequency of synchrotron radiation depends on the square of the WIMP mass \citep{rybicki86}. The SKA has also been suggested as a promising future instrument for DM searches \citep{colafrancesco15}.

\begin{acknowledgements}

We thank the anonymous referee for a constructive report which helped to improve the paper.

This work is funded by the Deutsche Forschungsgemeinschaft (DFG, German Research Foundation) under Germany's Excellence Strategy -- EXC-2121 "Quantum Universe" -- 390833306.

LOFAR \citep{vanhaarlem13} is the Low Frequency Array designed and constructed by ASTRON. It has observing, data processing, and data storage facilities in several countries, which are owned by various parties (each with their own funding sources), and that are collectively operated by the ILT foundation under a joint scientific policy. The ILT resources have benefited from the following recent major funding sources: CNRS-INSU, Observatoire de Paris and Université d'Orléans, France; BMBF, MIWF-NRW, MPG, Germany; Science Foundation Ireland (SFI), Department of Business, Enterprise and Innovation (DBEI), Ireland; NWO, The Netherlands; The Science and Technology Facilities Council, UK; Ministry of Science and Higher Education, Poland; The Istituto Nazionale di Astrofisica (INAF), Italy.

This research made use of the Dutch national e-infrastructure with support of the SURF Cooperative (e-infra 180169) and the LOFAR e-infra group. The Jülich LOFAR Long Term Archive and the German LOFAR network are both coordinated and operated by the Jülich Supercomputing Centre (JSC), and computing resources on the supercomputer JUWELS at JSC were provided by the Gauss Centre for Supercomputing e.V. (grant CHTB00) through the John von Neumann Institute for Computing (NIC). 

This research made use of the University of Hertfordshire high-performance computing facility and the LOFAR-UK computing facility located at the University of Hertfordshire and supported by STFC [ST/P000096/1], and of the Italian LOFAR IT computing infrastructure supported and operated by INAF, and by the Physics Department of Turin university (under an agreement with Consorzio Interuniversitario per la Fisica Spaziale) at the C3S Supercomputing Centre, Italy.

DJB acknowledges funding from the German Science Foundation DFG, via the Collaborative Research Center SFB1491 "Cosmic Interacting Matters - From Source to Signal".

\end{acknowledgements}

\bibliographystyle{aa}
\bibliography{references.bib}
\appendix

\onecolumn\section{Results for the individual galaxies}
\label{a:individual_galaxies}

\subsection{Radial intensity profiles}

\begin{figure}[!htb]
    \centering
    \vspace{-0.1cm}
    \begin{subfigure}[b]{0.45\textwidth}
        \centering
        \includegraphics[width=\linewidth]{CVNI_profile.pdf}
        \caption{}
        \label{fig:CVNI_RP_appendix}
    \end{subfigure}%
    \begin{subfigure}[b]{0.45\textwidth}
        \centering
        \includegraphics[width=\linewidth]{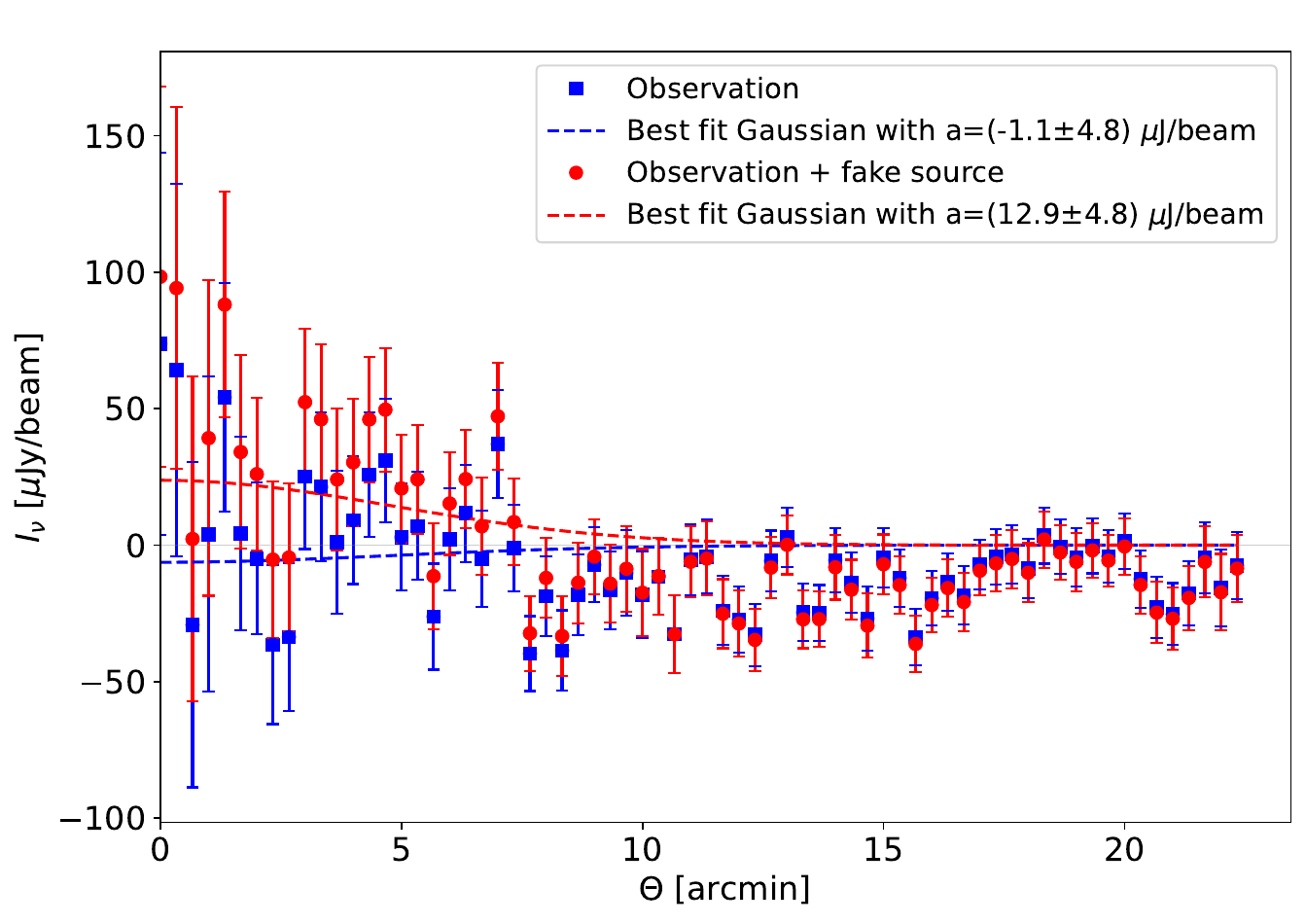}
        \caption{}
        \label{fig:USMI_RP_appendix}
    \end{subfigure}

    \vspace{0.2cm}
    \begin{subfigure}[b]{0.45\textwidth}
        \centering
        \includegraphics[width=\linewidth]{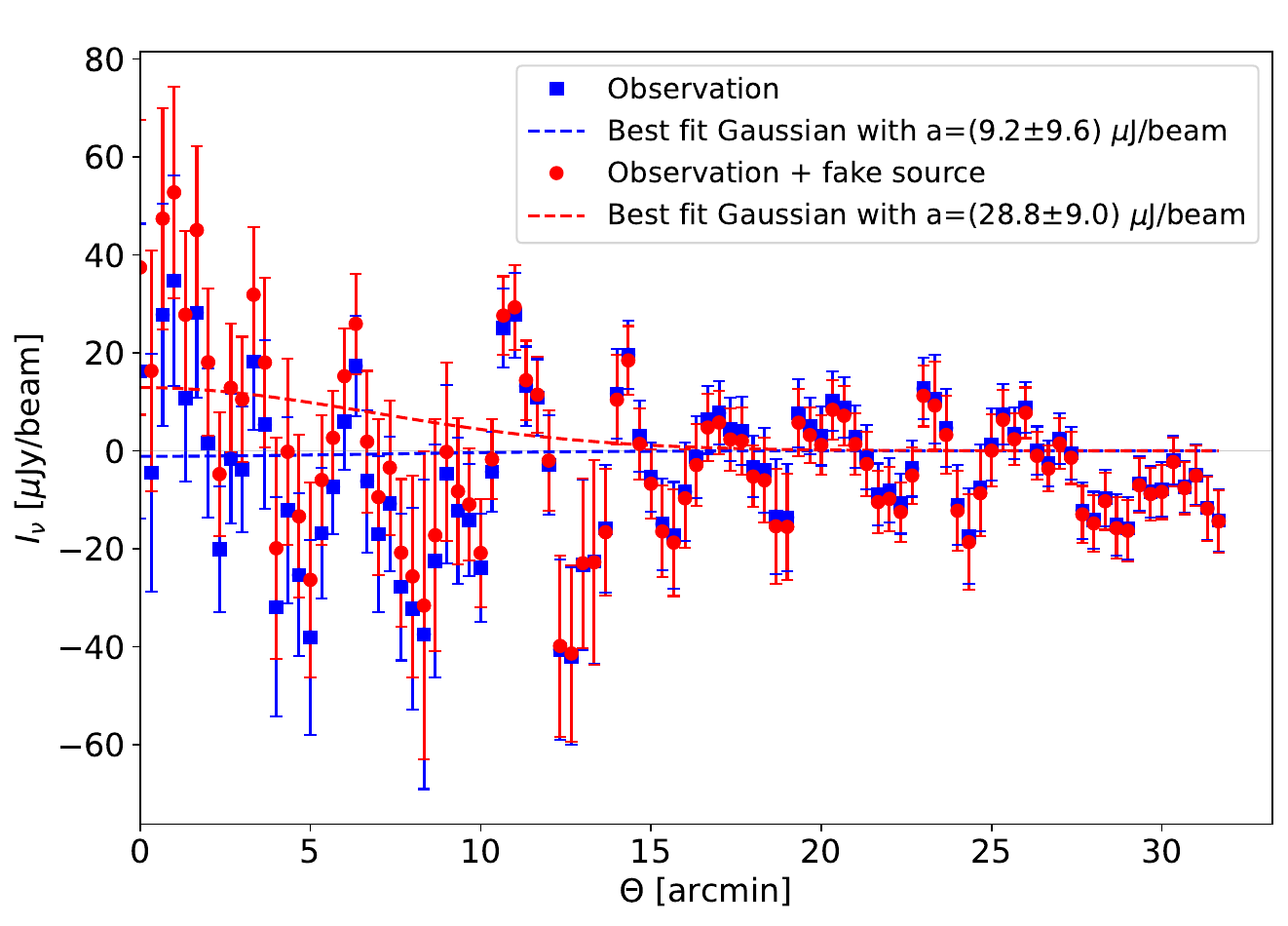}
        \caption{}
        \label{fig:USMII_RP_appendix}
    \end{subfigure}%
    \begin{subfigure}[b]{0.45\textwidth}
        \centering
        \includegraphics[width=\linewidth]{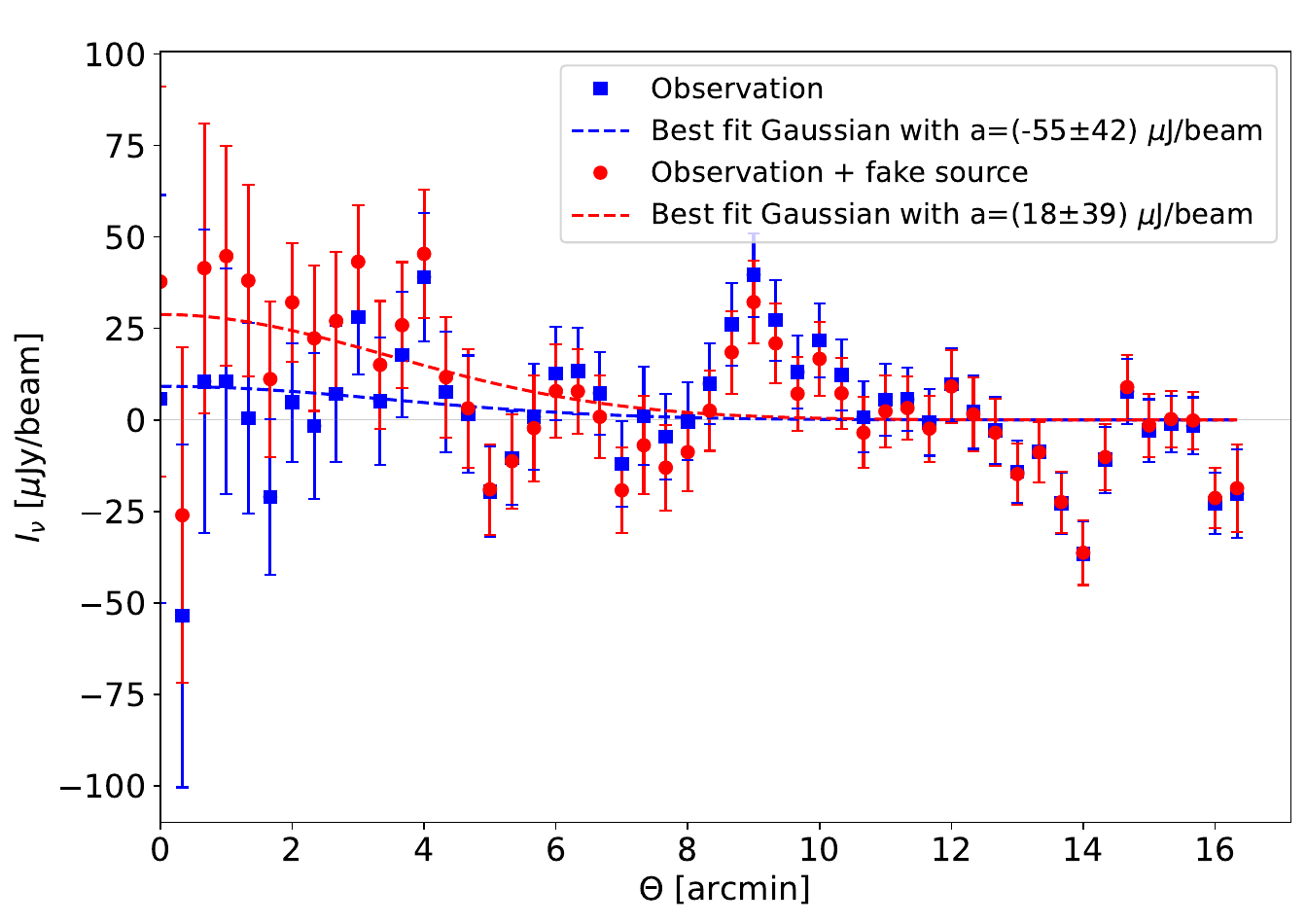}
        \caption{}
        \label{fig:USMN_RP_appendix}
    \end{subfigure}

    \vspace{0.2cm}
    \begin{subfigure}[b]{0.45\textwidth}
        \centering
        \includegraphics[width=\linewidth]{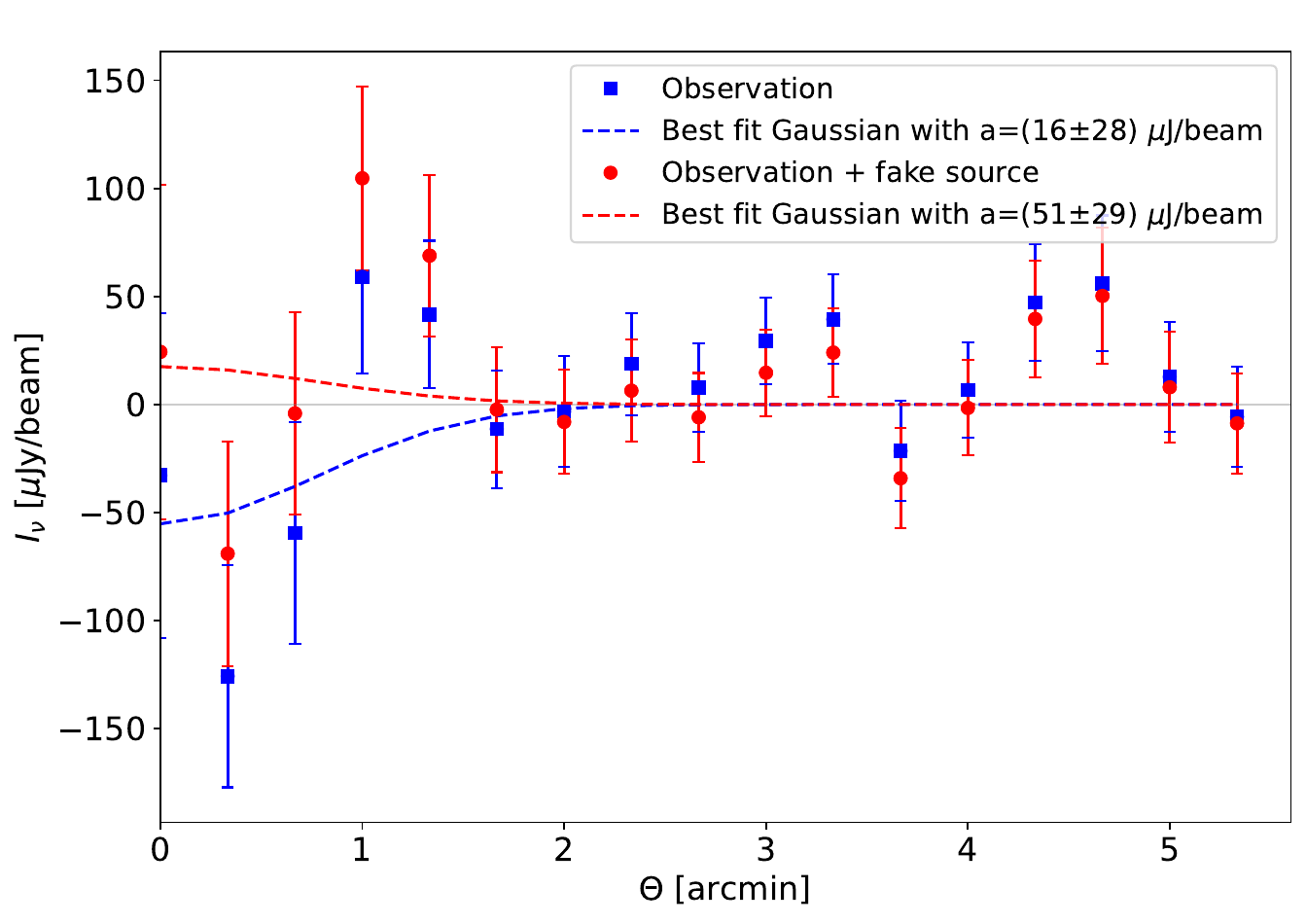}
        \caption{}
        \label{fig:WLMN_RP_appendix}
    \end{subfigure}%
    \begin{subfigure}[b]{0.45\textwidth}
        \centering
        \includegraphics[width=\linewidth]{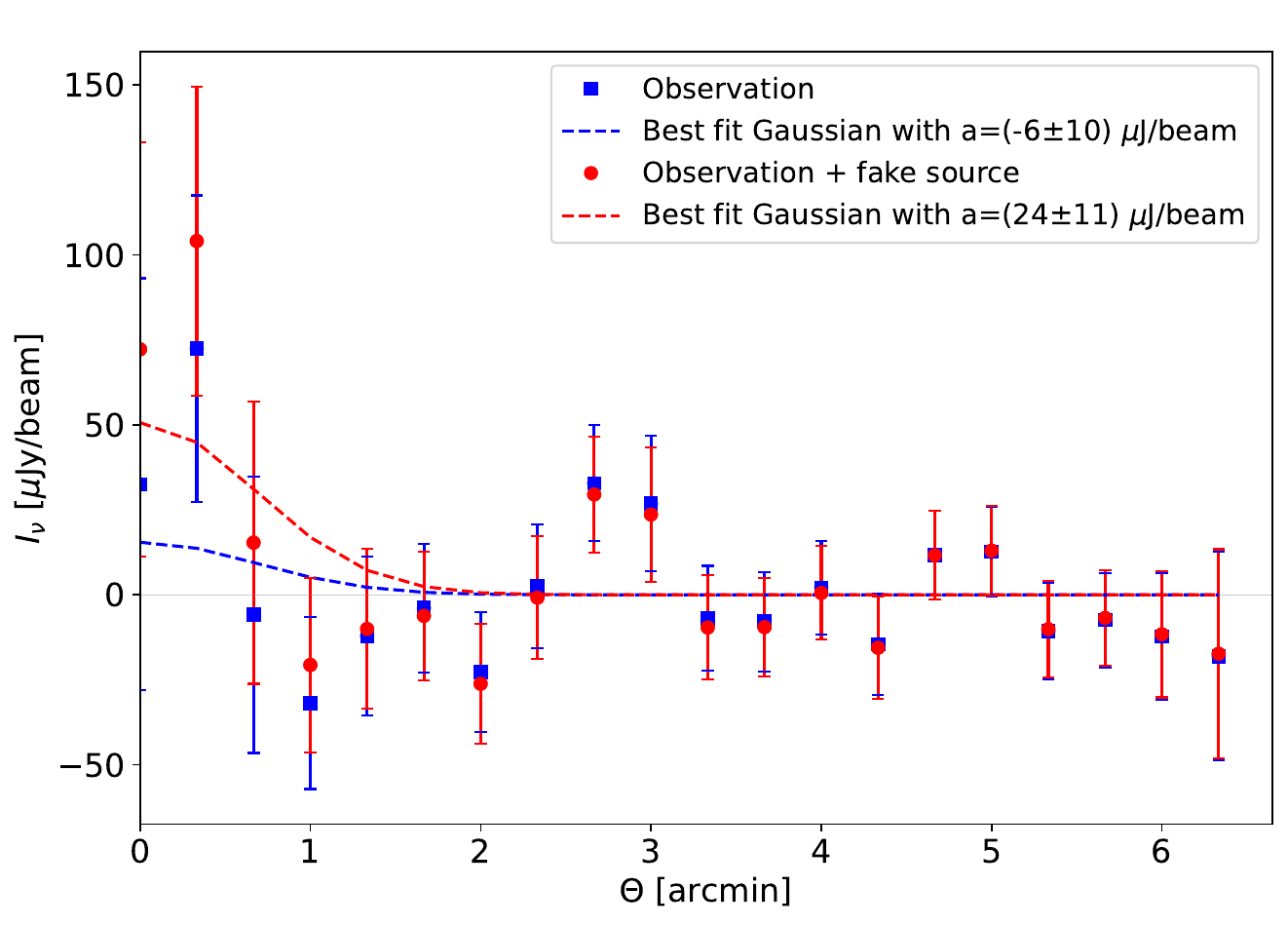}
        \caption{}
        \label{fig:CVNII_RP_appendix}
    \end{subfigure}
    \label{fig:radial_profiles_appendix}
    \caption{Radial intensity profiles from each of the six dSph galaxies: Canes Venatici I (panel a), Ursa Major I (b),  Ursa Major II (c), Ursa Minor (d), Willman I (e), and Canes Venatici II (f). Blue indicates purely observational data, and red indicates data with an additional flux density from a fake source. The individual flux densities are listed in Table \ref{tab:fake_source_fits}. The dashed lines show the best-fitting Gaussians with a FWHM of $r_\star$. The half-light radii are listed in Table \ref{tab:galaxy_samples}.}
\end{figure}

\vspace{1.5cm}
\subsection{Fitting plots}
\vspace{-0.2cm}
\begin{figure}[!htb]
    \centering
    \vspace{-0.4cm}
    \begin{subfigure}[b]{0.49\textwidth}
        \centering
        \includegraphics[width=\linewidth]{CVNI_FWHM.pdf}
        \caption{}
        \label{fig:CVNI_FWHM_appendix}
    \end{subfigure}
    \begin{subfigure}[b]{0.49\textwidth}
        \centering
        \includegraphics[width=\linewidth]{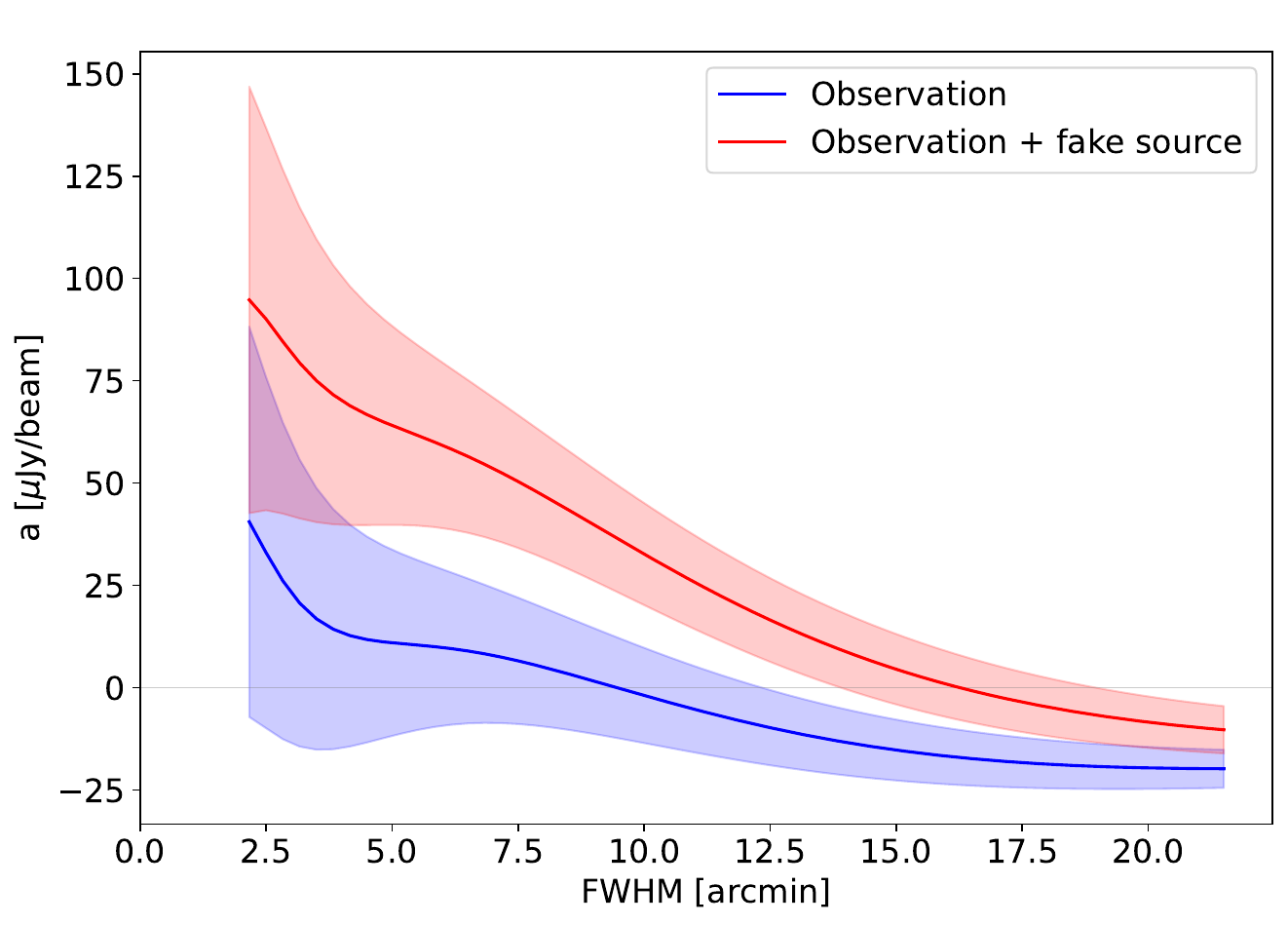}
        \caption{}
        \label{fig:USMI_FWHM_appendix}
    \end{subfigure}
    
    \vspace{0.2cm}
    \begin{subfigure}[b]{0.5\textwidth}
        \centering
        \includegraphics[width=\linewidth]{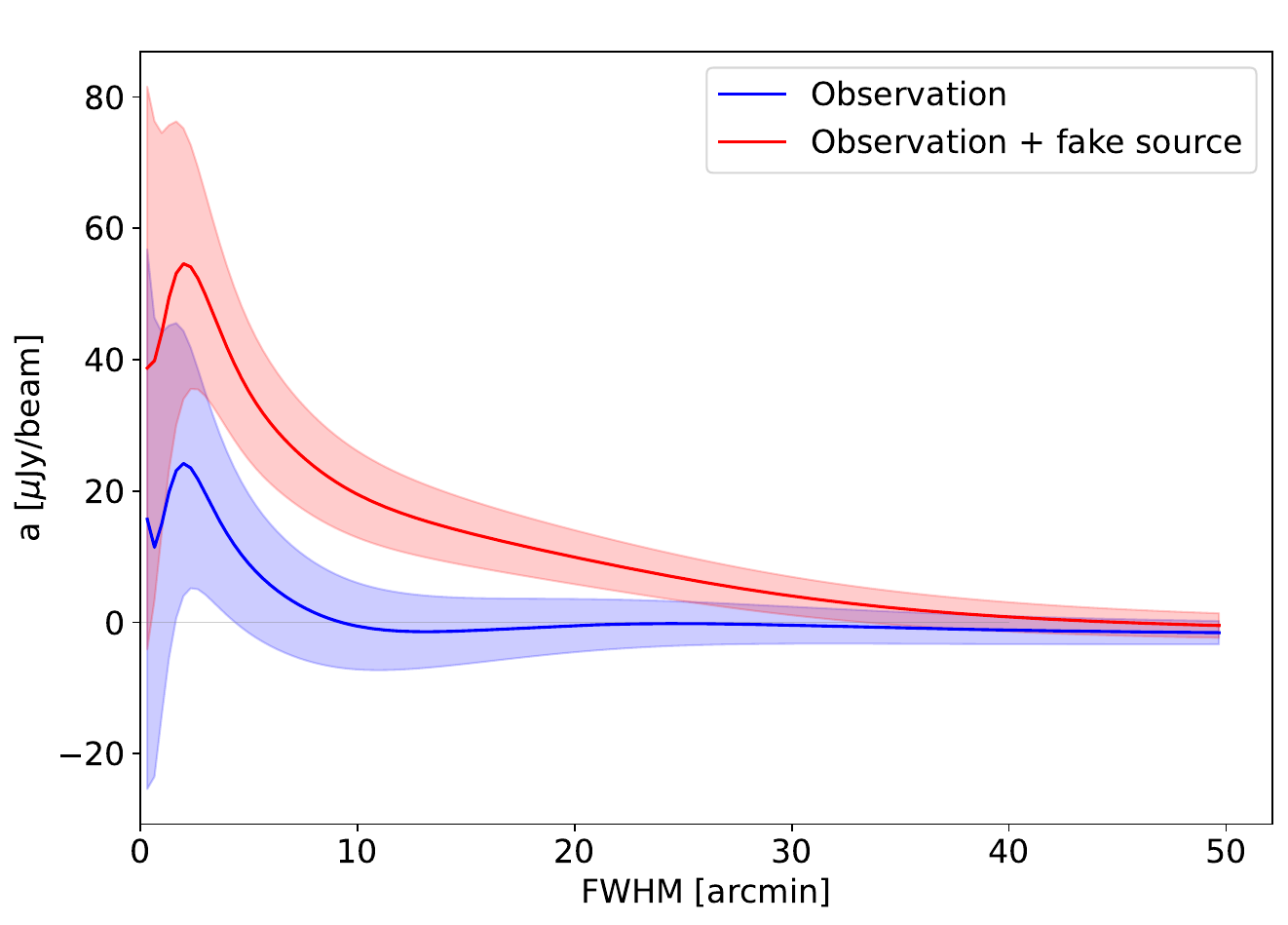}
        \caption{}
        \label{fig:USMII_FWHM_appendix}
    \end{subfigure}%
    \begin{subfigure}[b]{0.5\textwidth}
        \centering
        \includegraphics[width=\linewidth]{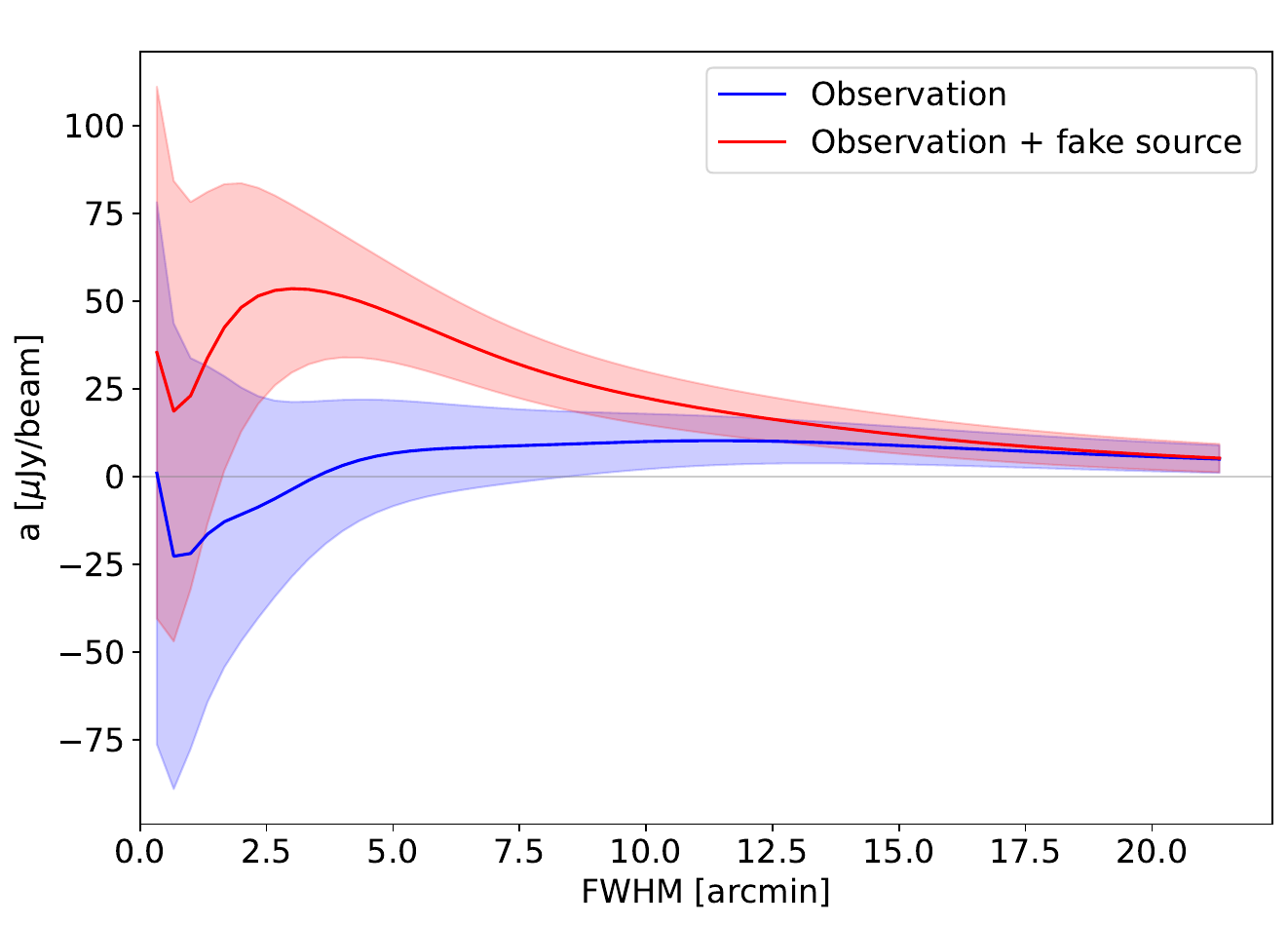}
        \caption{}
        \label{fig:USMN_FWHM_appendix}
    \end{subfigure}
    
    \vspace{0.2cm}
    \begin{subfigure}[b]{0.5\textwidth}
        \centering
        \includegraphics[width=\linewidth]{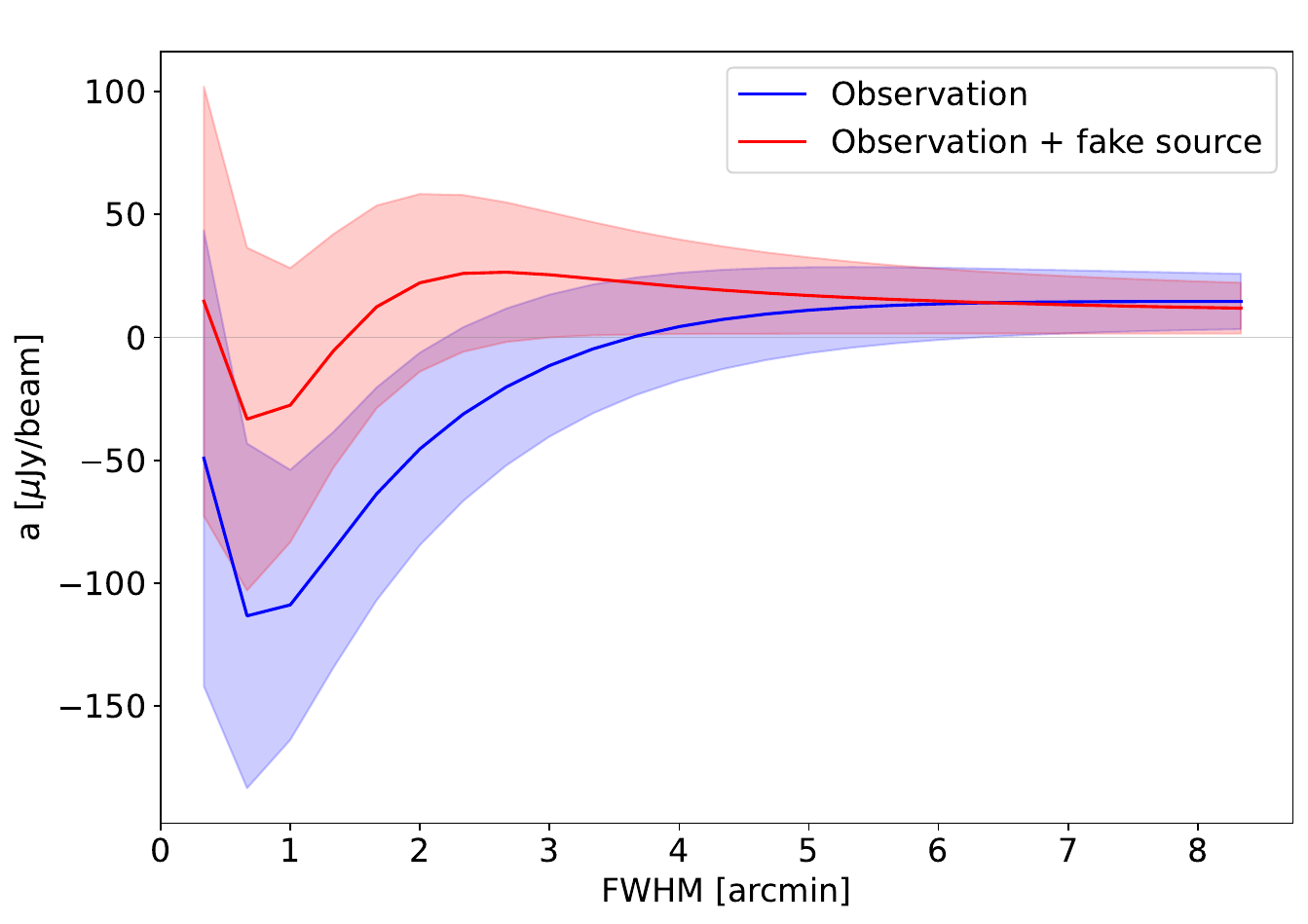}
        \caption{}
        \label{fig:WLMN_FWHM_appendix}
    \end{subfigure}%
    \begin{subfigure}[b]{0.5\textwidth}
        \centering
        \includegraphics[width=\linewidth]{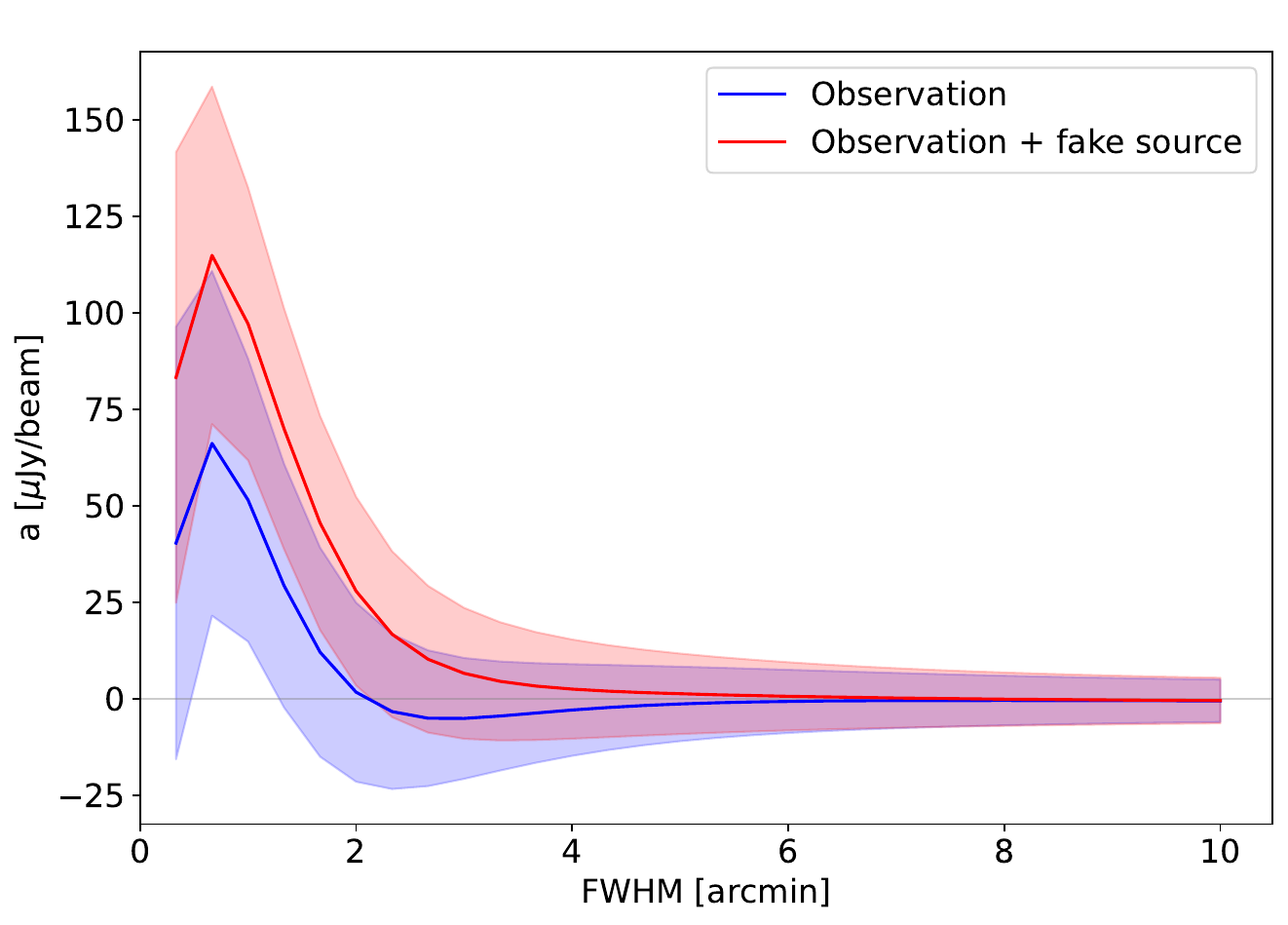}
        \caption{}
        \label{fig:CVNII_FWHM_appendix}
    \end{subfigure}
    \label{fig:fitting_plots_appendix}
    \caption{ Best-fitting Gaussian amplitudes for the radial intensity profiles in Figs. \ref{fig:CVNI_RP_appendix}- \ref{fig:CVNII_RP_appendix}: Canes Venatici I (panel a), Ursa Major I (b), Ursa Major II (c), Ursa Minor (d), Willman I (e), and Canes Venatici II (f). Blue indicates purely observational data, and red indicates data with an additional fake source. The added individual flux densities are listed in Table \ref{tab:fake_source_fits}. The shaded areas are $1\sigma$ intervals.}
\end{figure}
\newpage
\section{Reimaging parameters}
\label{a:re-calibration_parameters}
\begin{table}[ht]
    \caption{Parameters used to reimage each of the six galaxies.}
    \centering
    \begin{tabular}{lcccc}
     \hline \hline
        dSph & Lower & Robustness &  Wavelet  & Baseline-av. \\
              & $uv$-cut [$\lambda$] &  & scale & factor \\
        \hline
        \noalign{\smallskip}
        CVnI & 160 & -0.2 & 564 & 8.5221  \\
        UMaI & 60 & -0.2 & 319 & 5.6113 \\
        UMaII & 60 & -0.2 & 149 & 7.0936  \\
        UMi & 160 & -0.2 & 181 & 7.0025  \\
        WilI & 400 & -0.2 & 21 & 5.2477 \\
        CVnII & 400 & -0.2 & 74 & 10.3342 \\
        \hline
    \end{tabular}
    \label{tab:re-calibration_parameters}
\end{table}

\section{Stacked limits excluding the galaxy Willman I}
\label{a:noWLMI}

\begin{figure*}[h!]
    \centering
    \includegraphics[width=0.7\hsize]{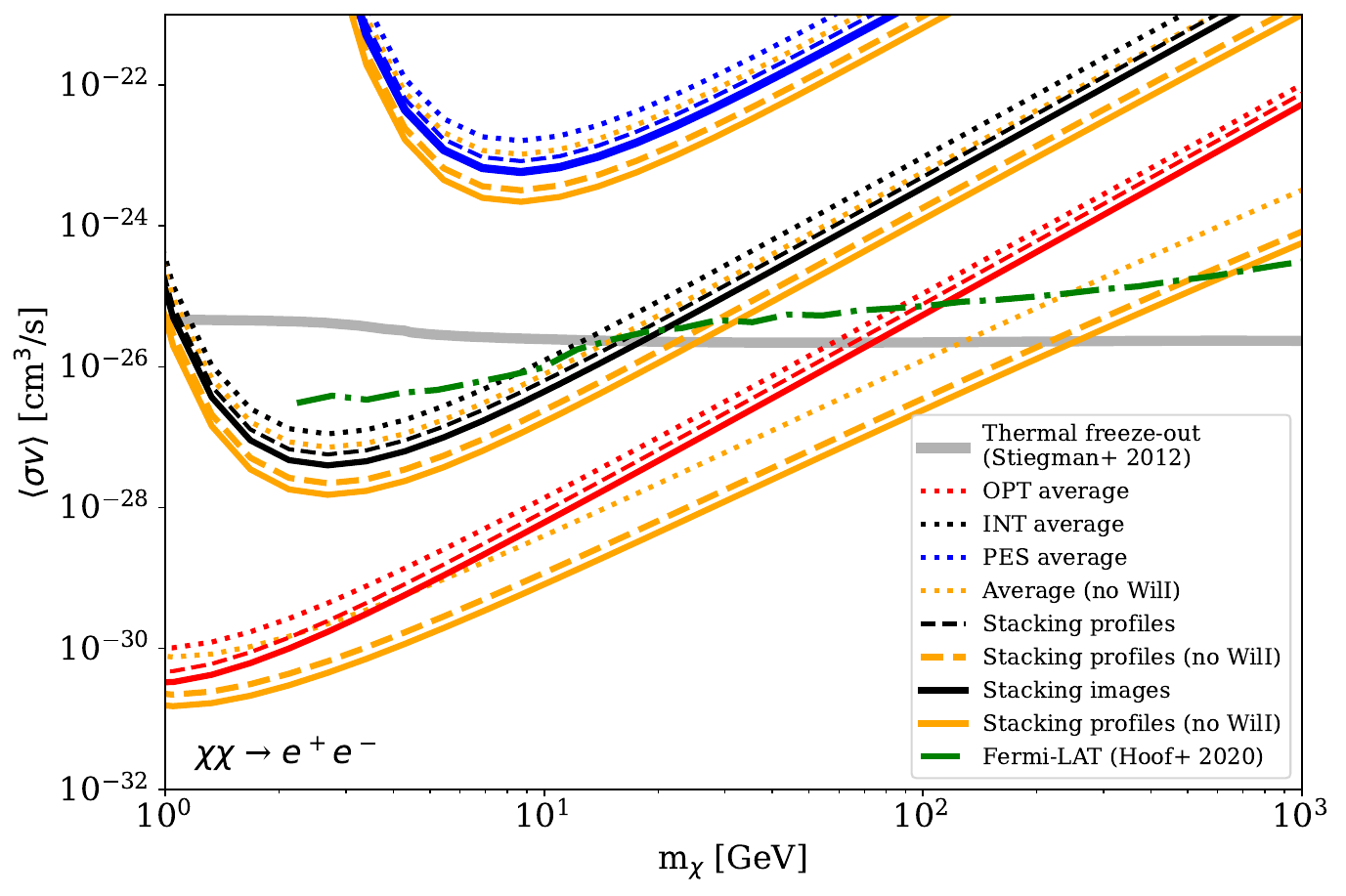}
    \caption{Upper limits on the WIMP annihilation cross section from stacking without the galaxy Willman I (in orange) compared to the results of stacking all galaxies (colors have the same meaning as in Fig. \ref{fig:stacked}). }
    \label{fig:stacked_noWLMI}
\end{figure*}

\end{document}